\documentclass[a4paper]{article}

\usepackage[margin=2cm]{geometry} 

\usepackage{amsmath}
\usepackage{amsthm}
\usepackage{amssymb}
\usepackage{mathtools}

\usepackage[utf8]{inputenc}
\usepackage{hyperref}

\usepackage[natbib=true, style=alphabetic]{biblatex}
\usepackage{authblk}

\usepackage{siunitx}
\DeclareSIUnit[]\molar{M}
\DeclareSIUnit[]\SHE{V_{SHE}}
\DeclareSIUnit[]\RHE{V_{RHE}}
\DeclareSIUnit[]\decade{dec}

\usepackage{graphicx, color}

\usepackage{booktabs}

\usepackage{enumerate}
\usepackage{enumitem}

\usepackage[version=3]{mhchem}

\title{
How Metal/Insulator Interfaces Enable the Enhancement of the Hydrogen Evolution Reaction Kinetics in Two Ways}

\author[1]{Thomas L. Maier\footnote{thomas.maier@ph.tum.de}}
\author[1]{Lucas B.T. de Kam}
\author[2]{Matthias Golibrzuch}
\author[1]{Tina Angerer}
\author[2]{Markus Becherer}
\author[1]{Katharina Krischer\footnote{krischer@tum.de}}

\affil[1]{Nonequilibrium Chemical Physics, Department of Physics, Technical University of Munich, 85748 Garching, Germany}
\affil[2]{Nano and Quantum Sensors, Department of Electrical and Computer Engineering, Technical University of Munich, 80333 München, Germany}

\date{ \today }

\begin{document}
\maketitle

\begin{abstract}
	Laterally nanostructured surfaces give rise to a new dimension of understanding and improving electrochemical reactions. In this study, we present a peculiar mechanism appearing at a metal/insulator interface, which can significantly enhance the Hydrogen Evolution Reaction (HER) from water reduction by altering the local reaction conditions in two ways: facilitated adsorption of hydrogen on the metal catalyst surface and improved transfer of ions through the double layer. The mechanism is uncovered using electrodes consisting of well-defined nanometer-sized metal arrays (Au, Cu, Pt) embedded in an insulator layer (silicon nitride), varying various parameters of both the electrode  (size of the metal patches, catalyst material) and the electrolyte (cationic species, cation concentration, pH). In addition, simulations of the electrochemical double layer are carried out, which support the elaborated mechanism. Knowledge of this mechanism will enable new design principles for novel composite electrocatalytic systems.

\end{abstract}


\section{Introduction}\label{sec:intro}

Nanostructured or non-homogeneous surfaces often show a strongly altered electrocatalytic behavior compared to homogeneous surfaces at electrochemical interfaces \cite{Subbaraman2011, Danilovic2012, Maier2020, Wu2021}. Changes in adsorption of reaction intermediates \cite{Chen2017, Xue2018}, changes in the reaction mechanism \cite{Subbaraman2011, Danilovic2012} or changes in the electrochemical double layer \cite{LedezmaYanez2017, Goyal2021, Monteiro2021} are commonly mentioned causes. Thus, nanostructuring provides an additional dimension for understanding fundamental processes at electrochemical interfaces and another way to optimize the kinetics of the desired reaction.

Let us consider the Hydrogen Evolution Reaction (HER) as an exemplary reaction. The HER is a two-electron transfer reaction consisting of electrochemical (Volmer, Heyrovsky) or chemical (Tafel) elementary steps. In alkaline media, these steps are given by:
\begin{alignat}{3}
	&\text{Volmer:}\qquad &&\ce{H2O + e- + ^* &&-> H_{ad} + OH-}\label{eq:Volmer} \\
	&\text{Heyrovsky:}\qquad &&\ce{H2O + H_{ad} + e^- &&-> H2 + OH- + ^*}\label{eq:Heyrovsky} \\
	&\text{Tafel:}\qquad &&\ce{2H_{ad} &&-> H2 + 2 ^*}\label{eq:Tafel}
\end{alignat}
Here $^*$ denotes a free adsorption site. There are multiple examples in the literature showing that HER kinetics and its parameter dependencies can be altered significantly on non-homogeneous interfaces \cite{Esposito2013, Strmcnik2016, Filser2018, Maier2020, Wu2021} compared to bare metal surfaces.
For example, Markovic et al. \cite{Danilovic2012} showed that the coverage of various metal surfaces by metal-oxide particles (in particular $\ce{Ni(OH)2}$ nanoparticles) can enhance the HER kinetics significantly. Due to the surface modification, the activity of the metal surfaces in alkaline medium approaches their activity in acidic medium. The authors explained the enhancement by the appearance of a bifunctional mechanism at the metal/particle interface promoting the water dissociation step. Koper and coworkers \cite{LedezmaYanez2017} proposed that the introduction of $\ce{Ni(OH)2}$ particles promotes the transport of hydroxide into the electrolyte bulk, as this changes the compactness of the double layer and, thus, lowers the energy barrier necessary for the reorganization of the interfacial water layer. However, the actual mechanistic reason responsible for the observed enhancement is still debated in the literature \cite{Subbaraman2011, Danilovic2012, LedezmaYanez2017, Sarabia2018, Rebollar2018}.

In a previous study \cite{Maier2020} we showed that
an insulator/metal interface provides a further means to enhance an electrochemical reaction, using the HER from water as an example. Here we show that the cause of this unique amplification is due to two interacting mechanisms, effecting adsorption and double layer properties of the electrified interface. Specifically, we investigate the HER on nanostructured electrodes consisting of a well-defined array of metal structures embedded in a silicon-based insulator layer (silicon nitride). By varying the properties of the electrode (array geometry and catalyst material) and electrolyte (pH, cation species, and cation concentration), we uncover the unique role of the metal/insulator interface in the observed increase in reaction rate. 
Using simulations of the electrochemical double layer, we calculate the fraction of protonated silanol groups on the insulator surface and the pressure across the metal and insulator regions, which results from electrostatic forces and can be considered a measure of the stiffness of the double layer. Based on these quantities, we derive a qualitative model for the current density that captures all experimentally observed trends at different electrode and electrolyte parameters, thus supporting the proposed mechanism.

\section{Results}\label{sec:results}

\begin{figure}[h!]
	\centering
	\includegraphics[width=0.49\textwidth]{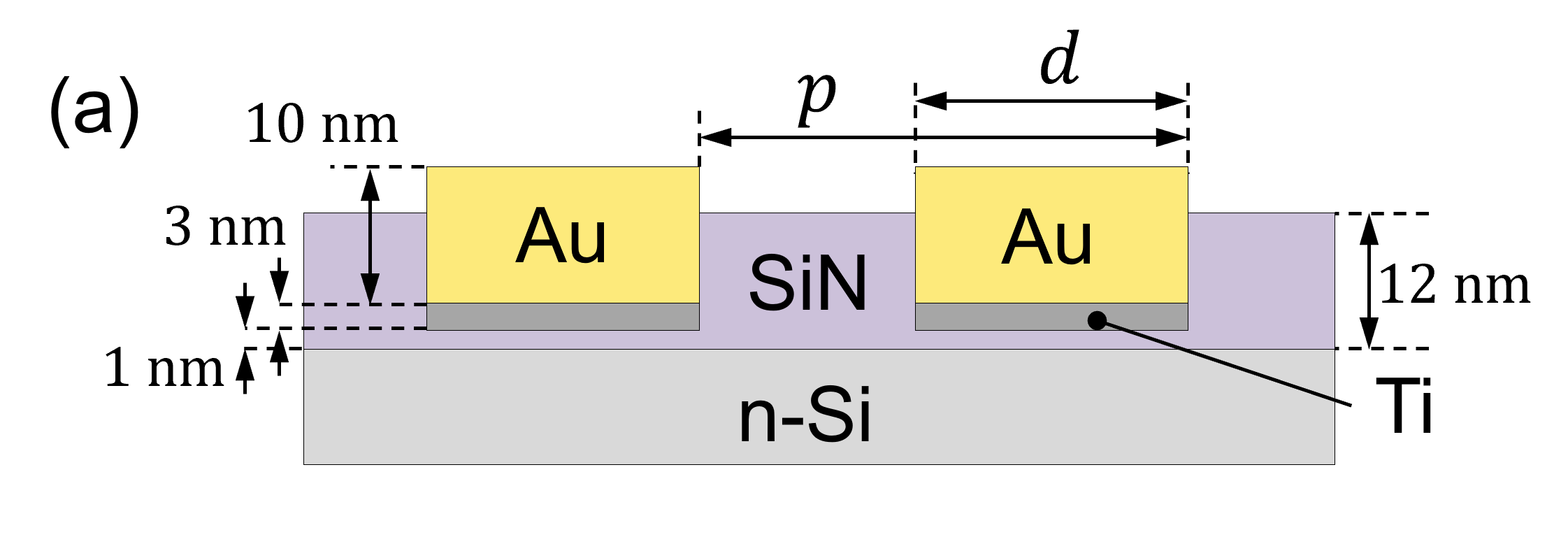}
	\includegraphics[width=0.4\textwidth]{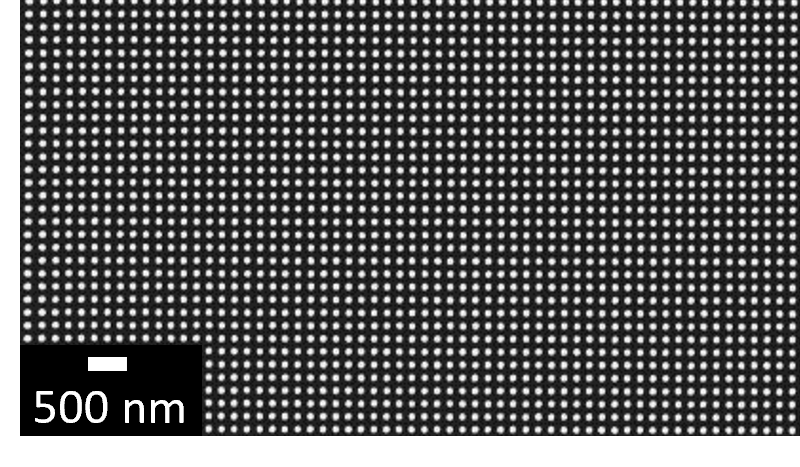}
	\caption{(a) Schematic intersection of the investigated nanostructured electrodes. The size $d$ of individual structures comprises $\qty{1400}{\nano\meter}$, $\qty{350}{\nano\meter}$, $\qty{200}{\nano\meter}$, and $\qty{75}{\nano\meter}$ (smallest structures). The pitch ratio $d/p$ of the arrays is rather similar for all electrodes investigated, which gives a rather similar coverage for different $d$. A table listing the actual geometric values of the arrays is given in the appendix \ref{app:geometric-properties}. (b) Exemplary SEM image of a nanostructure array with structure size $d= \qty{75}{\nano\meter}$. White: Gold structures, Dark: silicon nitride surface surrounding the gold structures.}
	\label{fig:scheme-electrode}
\end{figure}

We analyze silicon-based electrodes, the surfaces of which are covered by arrays of metal nano-islands (at first Au) embedded in an insulator (silicon nitride, SiN) layer, resulting in a laterally structured surface that is exposed to the electrolyte.
A schematic intersection of the nanostructured electrode can be found in Figure \ref{fig:scheme-electrode}(a). Important geometric quantities are labeled in the Figure. A table listing the geometric parameters of all nanostructure arrays used is given in the appendix \ref{app:geometric-properties}. An exemplary SEM image depicting the top view of an array of nanostructures with diameter 75 nm is shown in Figure \ref{fig:scheme-electrode}(b).

The insulator layer based on silicon nitride solves the problem of stability in alkaline environment compared to a silicon oxide based insulating layer, which were considered in a previous study \cite{Maier2020}. The silicon nitride based systems give, as shown below, similar kinetic results as the silicon oxide based systems. This further shows the generality of the investigated effect for various (silicon based) insulator layers.

\subsection{Increased HER rate at the metal/insulator interface}

\begin{figure}[h!]
	\centering
	\includegraphics[width=0.49\textwidth]{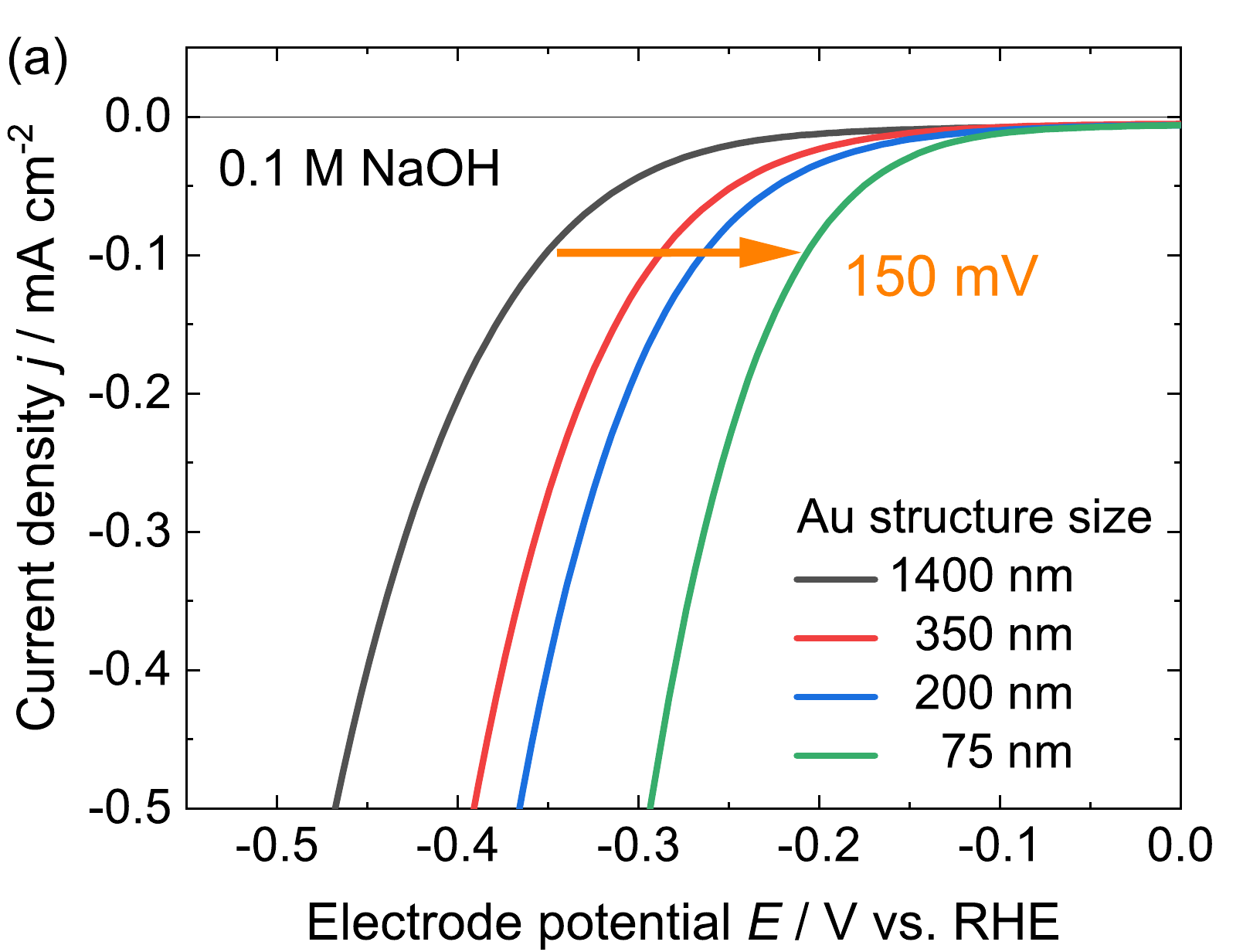}\\
	\vspace{0.2cm}
	\includegraphics[width=0.35\textwidth]{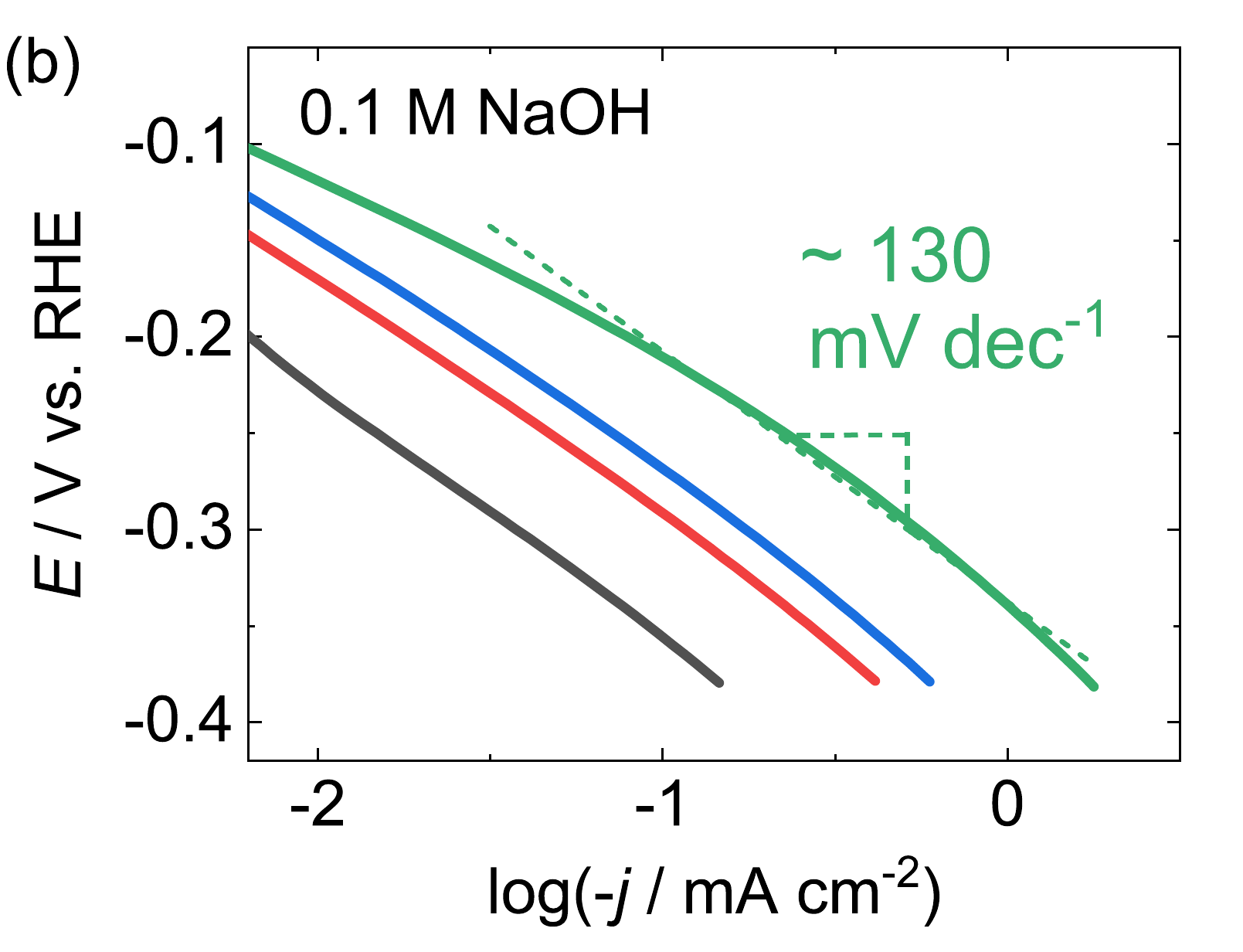}
	\includegraphics[width=0.35\textwidth]{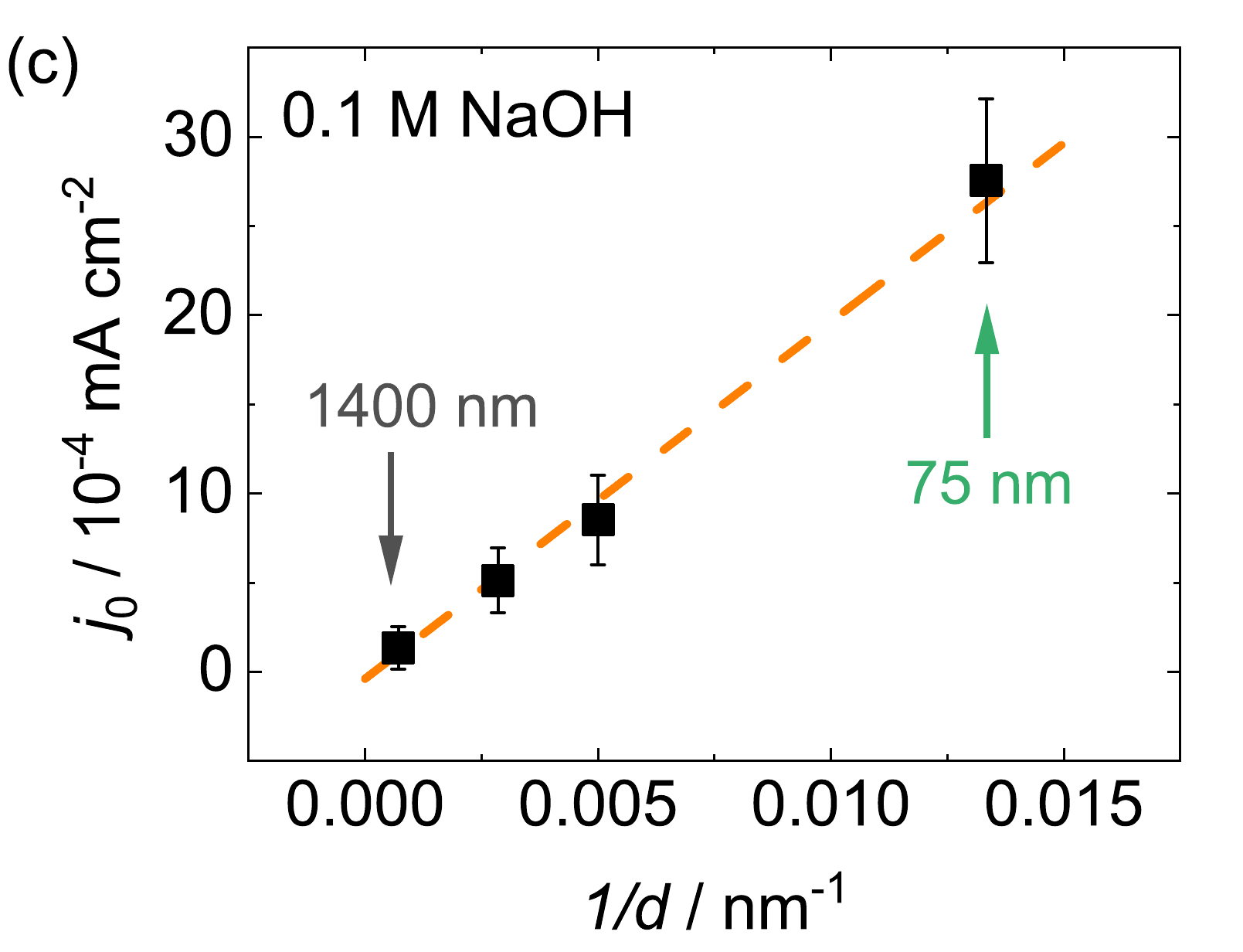}
	\caption{(a) LSVs of the Au-based nanostructure arrays in the HER region in $\qty{0.1}{\molar}~\ce{NaOH}$/Ar sat.'d. Scan rate: $\qty{50}{\milli\volt\per\second}$. (b) Tafel plot of the data after subtraction of the non-faradaic current at $\qty{0}{\RHE}$. All curves proceed quite parallel with a slope between $120-\qty{130}{\milli\volt\per\decade}$. (c) Determined exchange current density $j_0$ plotted over the inverse of the structure size $d$. The dotted orange line is a guide to the eye and indicates the linear dependence.}
	\label{fig:size-series_si3n4}
\end{figure}

By changing the size of the structures while keeping the metal coverage constant, the total length of the metal/insulator interface can be tuned. A detailed description can be found in a previous publication \cite{Maier2020}. Figure \ref{fig:size-series_si3n4} shows the dependence of the HER activity of the electrodes in $\qty{0.1}{\molar}~\ce{NaOH}$/Ar sat.'d on the size of the individual Au structures arranged in an array. It becomes evident that decreasing the structure size increases the current density of HER. This manifests itself in an overpotential decrease by approx. $\qty{150}{\milli\volt}$ at $-\qty{0.1}{\milli\ampere\per\square\centi\meter}$ between the largest ($d=\qty{1400}{\nano\meter}$) and the smallest ($d=\qty{75}{\nano\meter}$) structures investigated, cf. Figure \ref{fig:size-series_si3n4}(a).

The plot of the logarithmic current as a function of the electrode potential (Tafel plot), cf. Figure \ref{fig:size-series_si3n4}(b), shows a rather constant, i.e. potential independent, slope, which is similar for all Au structure sizes. Its value is $\approx\qty{130}{\milli\volt\per\decade}$. A slope around $\qty{120}{\milli\volt\per\decade}$ indicates that the Volmer step is the rate-determining-step \cite{Ohmori1992, Rheinlaender2014, LedezmaYanez2017, Zheng2018, Goyal2021}. The rate in the considered current density range is not determined by transport, as shown in  appendix \ref{app:rotation-speed}.

The exchange current densities, as obtained from the Tafel plot, scale linearly with the inverse of the structure sizes $d$ (Figure \ref{fig:size-series_si3n4}(c)). This linear dependence implies that the improved HER activity observed for smaller Au structures stems from a higher reactivity of HER at the boundary of the metal structures (i.e. at the metal/insulator interface) compared to the centre of the metal structures (i.e. the 'metal-bulk') \cite{Maier2020}.

In the following, we  present parameter studies for three different model systems that reveal the cause of the observed amplification: A continuous gold layer electrode (Au cont., benchmark system), a gold array with rather large structures ($d=\qty{1400}{\nano\meter}$), i.e. a small total circumference of the gold/insulator interface, and a gold array with comparatively small structures ($d=\qty{75}{\nano\meter}$), i.e. a large gold/insulator interface length. Full CVs of the continuous Au layer and of the nanostructured electrode with smallest structures can be found in  appendix \ref{app:CVs-Au-film}. We change the pH value of the electrolyte, the cationic species present in the electrolyte, or the bulk cation concentration, which all alter the surface cation concentration and the double layer properties \cite{Goyal2021, Monteiro2021}. We will see that nanostructured electrodes show an altered dependence on the (surface) cation concentration compared to the one for continuous metal layers.

\subsection{Dependence on the electrolyte pH} \label{sec:results-ph}

\begin{figure}[h!]
	\centering
	\includegraphics[width=0.55\textwidth]{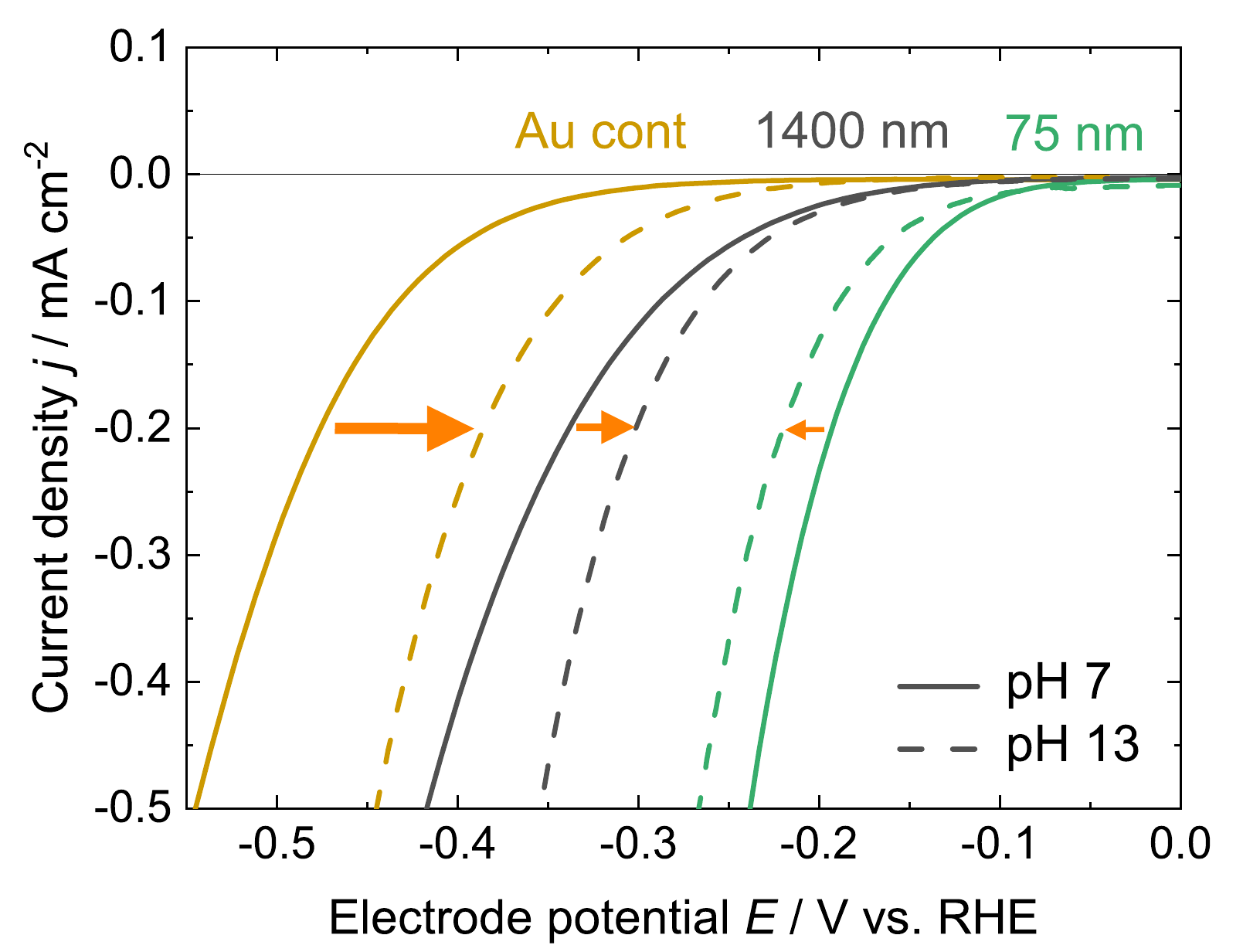}
	\caption{Comparison of HER activity for different Au-based electrode systems: Continuous layer electrode (yellow), nanostructured electrode with 1400 nm structure size (black), and  nanostructured electrode with 75 nm structure size (green). The graph compares LSVs conducted in neutral medium (solid lines, $\qty{0.18}{\molar}~\ce{K+}$ + $\qty{0.12}{\molar}~\ce{H2PO4-}/\ce{HPO4^2-}$, pH 7) and alkaline medium (dashed lines, $\qty{0.1}{\molar}~\ce{KOH}$, pH 13). Scan rate: $\qty{50}{\milli\volt\per\second}$. }
	\label{fig:pH-value}
\end{figure}

Figure \ref{fig:pH-value} shows LSV curves of the three mentioned electrode systems obtained in alkaline medium ($\qty{0.1}{\molar} ~\ce{KOH}$, pH 13) and in buffered neutral medium (pH 7, $\qty{0.18}{\molar}~\ce{KOH}$ + $\qty{0.12}{\molar}~\ce{H2PO4^{-}}/\ce{HPO4^{2-}}$). Note that the electrode potential is given vs. the RHE scale, in which the equilibrium potential for HER is the same in both electrolytes, but the point-of-zero-charge (PZC), which is approx. constant in SHE scale, differs. Consequently, the surface concentration of cations at the same potential in RHE scale is higher in alkaline electrolyte than it is in neutral electrolyt \cite{LedezmaYanez2017}.

We observe the following behavior for the three systems:
\begin{itemize}
	\item
	Continuous Au layer electrode (yellow lines): To achieve a certain reaction current, a larger overpotential is necessary at pH 7 than at pH 13 (see orange arrow in the figure here exemplary at $-\qty{0.2}{\milli\ampere\per\square\centi\meter}$). This behavior is in accordance with the behavior known from the literature for bare Au electrodes \cite{Goyal2021}. According to the literature, a higher concentration of cations in front of the electrode surface, as it is evident for alkaline medium compared to neutral medium in RHE scale, improves HER on Au electrodes, since the cations are supposed to stabilize the transition state of the water dissociation \cite{Goyal2021}.
	\item
	1400 nm electrode (black lines): In this system, the current-voltage curve measured in pH7 is again shifted to larger overpotentials compared to the one obtained in pH13, but considerably less than for the continuous Au layer electrode (cf. shorter orange arrow in the figure).
	\item
	75 nm electrode (green curves): In this system, both curves lie rather close to each other and even exhibit an opposite trend. To draw a given reaction current slightly less overpotential must be applied at pH 7 than at pH 13 (cf. orange arrow in negative direction).
\end{itemize}

In summary, the trend observed for a continuous Au layer electrode, namely that the HER activity increases when going from a neutral to an alkaline pH, is reversed for a system with a comparatively large contribution of the metal/insulator interface to the reaction current.

\subsection{Dependence on the cationic species} \label{sec:results-cat-species}

\begin{figure}[h!]
	\centering
	\includegraphics[width=0.99\textwidth]{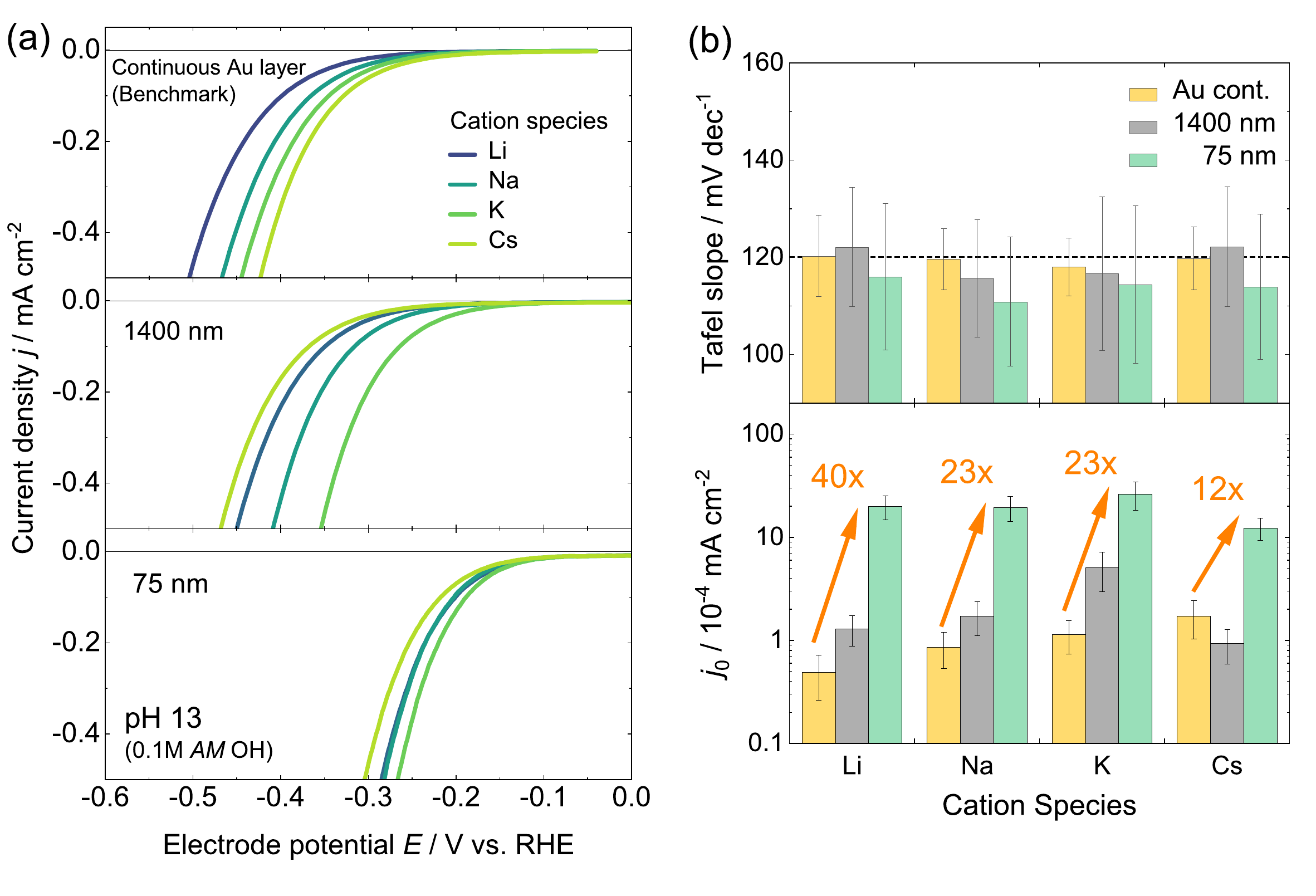}
	\caption{Dependence of the HER activity on the cationic species. (a) LSVs in $\qty{0.1}{\molar}~AM\ce{OH}$ electrolyte (pH 13) for different alkali metal cations ($AM = \ce{Li}, \ce{Na}, \ce{K}, \ce{Cs}$) and electrode systems: A continuous gold layer (top), a nanostrutured electrode with $\qty{1400}{\nano\meter}$ metal island diameter (middle), and a nanostructured electrode with $\qty{75}{\nano\meter}$ metal island diameter (bottom). (b) Determined Tafel slopes (top) and determined exchange current densities $j_0$ (bottom) of the three electrode systems for various cationic species.}
	\label{fig:cation-species}
\end{figure}

Next, we analyze the behavior of the HER rate of the three model systems when changing the cationic species present in alkaline electrolyte. Figure \ref{fig:cation-species}(a) shows LSVs of the three systems in alkaline medium with various alkali metal cation species in solution ($\qty{0.1}{\molar}~AM\ce{OH}$ with $AM= \ce{Li}, \ce{Na}, \ce{K}, \ce{Cs}$, pH 13 for all electrolytes).

We observe the following behavior:
\begin{itemize}
	\item
	Continuous layer electrode (Figure \ref{fig:cation-species}(a), top): The observed HER activity of the system depends on the nature of the cation species. We observe the following activity trend from highest to lowest: $\ce{Cs} > \ce{K} > \ce{Na} > \ce{Li}$. Thus, the activity  increases systematically down the periodic table, i.e. with decreasing degree of solvation of the ions \cite{Goyal2021}. This behavior is known from the literature \cite{Xue2018, Monteiro2021}. It is traced back to a varying net cation concentration at the electrode surface, which originates form different sizes of the solvation shells of cations. A higher net cation concentration at the electrode surface is supposed to stabilize the transition state of the rate determining Volmer step \cite{Goyal2021, Monteiro2021}.
	\item
	1400nm electrode (Figure \ref{fig:cation-species}(a), middle): Compared to the continuous Au layer system, the HER rate is higher for Li, Na, or K whereas the activity observed for Cs is slightly lower. Consequently, the order of activities changes from the highest to the lowest value and now reads: $\ce{K} > \ce{Na} > \ce{Li} > \ce{Cs}$.
	\item
	75nm electrode (Figure \ref{fig:cation-species}(a), bottom): The activity of this systems is much higher than the activity observed for the continuous layer system and the 1400 nm electrode. This is evident for all different cationic species investigated here. Furthermore, the activity order ranked from highest to lowest is now given by: $\ce{K} > \ce{Na} \approx \ce{Li} > \ce{Cs}$
\end{itemize}

Thus, the increase of the HER activity on nanostructured electrodes compared to bare metal surfaces appears to be present for all different cationic species investigated. A similar trend is also found in neutral electrolyte, which is shown in appendix \ref{app:cation-species-neutral}.

Figure \ref{fig:cation-species}(b) shows the Tafel analysis of the considered data. The upper panel displays the Tafel slopes for the three different systems and the various cationic species; the lower panel depicts the corresponding exchange current densities $j_0$. We can make the following observations:
\begin{itemize}
	\item
	Tafel slope: The slopes of all different electrode / electrolyte variations investigated show rather similar Tafel slopes with values around $\qty{120}{\milli\volt\per\decade}$ (cf. dashed line in the figure). Thus, in all systems and for all cationic species considered here, the HER rate appears to be consistently determined by the first electron transfer step, i.e., the Volmer step.
	This is well in accordance with results found in the literature for gold-based electrodes in alkaline media \cite{Goyal2021, Monteiro2021}.
	\item
	The comparison of the determined exchange current densities (Figure \ref{fig:cation-species}(b) bottom) shows that there is a huge difference of HER kinetics between the continuous Au layer system and the 75 nm electrode.
\end{itemize}

For the further discussion of the dependence of the HER activity on the cationic species we introduce a scaling factor $F_{AM}$, depending on the cation species $AM$. The factor gives the ratio between $j_0$ of the $\qty{75}{\nano\meter}$ electrode system and $j_0$ of the continuous layer system:
\begin{equation} \label{eq:scaling-factor}
	F_{AM} = \frac{j_{0,AM}(\qty{75}{\nano\meter})}{j_{0,AM}(\text{Au cont.})}
\end{equation}
The scaling factor normalizes the activity of an electrode with a rather large contribution of the metal/insulator to the data of a system without any metal/insulator border. Thus, $F_{AM}$ eliminates the intrinsic dependence of the HER rate on the cationic species present for the continuous metal surface. Instead, it quantifies the enhancement factor due to the structure size for each cationic species. We derive the following scaling factors from the data (Figure \ref{fig:cation-species}(b) bottom): $F_\mathrm{Li}\approx40$, $F_\mathrm{Na}\approx23$, $F_\mathrm{K}\approx23$, and $F_\mathrm{Cs}\approx12$ (see orange arrows in the Figure). $F$ appears to follow a systematic trend:
\begin{equation}
    F_\text{Li} > F_\text{Na} \approx F_\text{K} > F_\text{Cs}
\end{equation}
It decreases with decreasing degree of solvation. Thus, the HER activity at the Au/insulator interface shows the opposite trend with the size of the cationic species as the continuous Au layer.


\subsection{Dependence on the bulk cation concentration} 	\label{sec:results-cat-conc}

Next, we investigate the behavior of the HER activity of nanostructured electrodes on the (bulk) cation concentration. We analyze the activity in buffered neutral electrolytes for different (bulk) concentration of cations. Here we choose $\ce{K}$ as the cationic species present in solution. All electrolytes investigated exhibit a similar pH of 7. The composition of the electrolytes is given in appendix \ref{app:composition-cation-concentration}.

\begin{figure}[h!]
	\centering
	\includegraphics[width=0.55\textwidth]{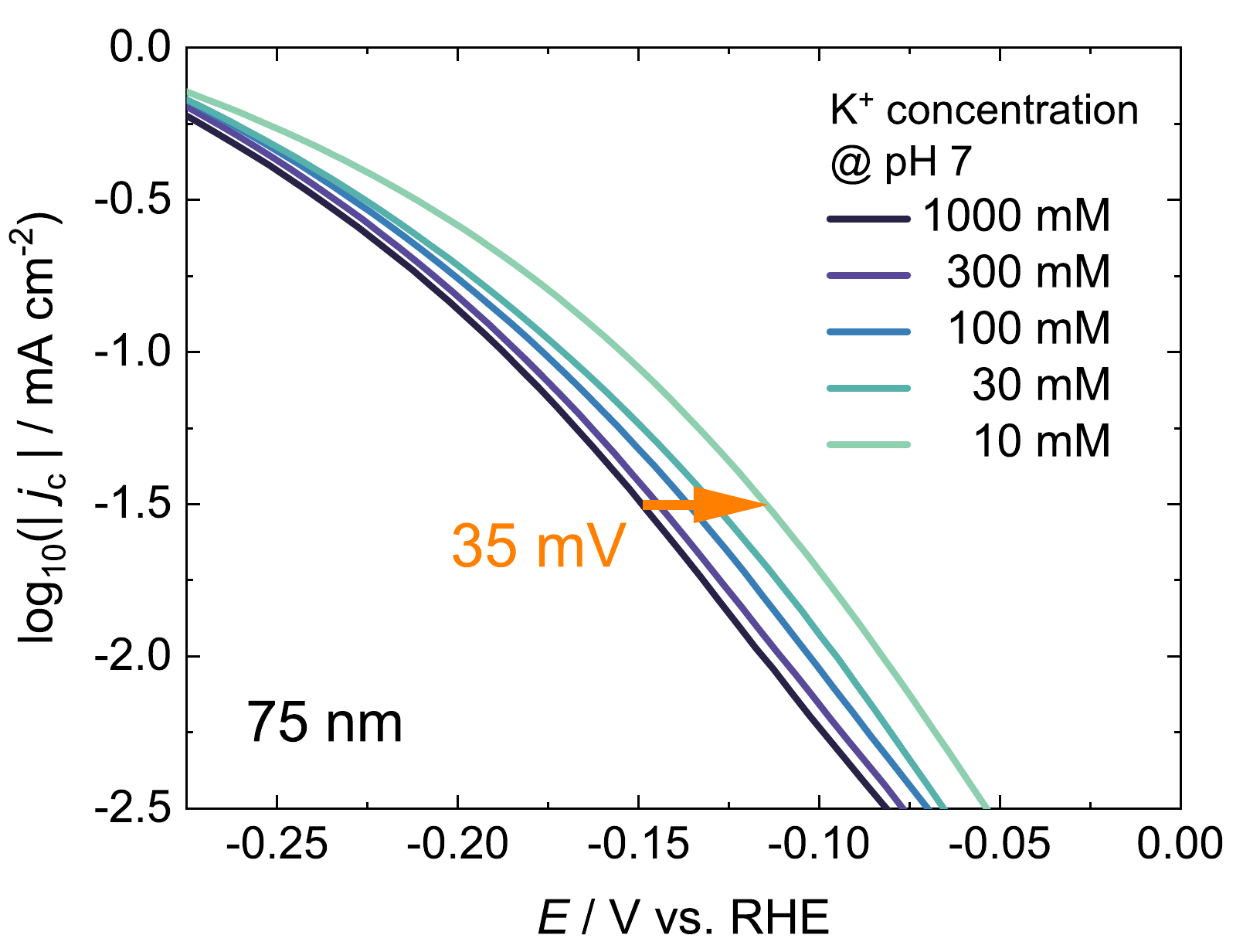}
	\caption{LSVs of a nanostructured electrode with $\qty{75}{\nano\meter}$ structure size in electrolytes with various cation $\ce{K+}$ concentration. All electrolyte have a similar pH value of 7. The data is IR-corrected. The exact composition of the electrolytes is given in the appendix \ref{app:composition-cation-concentration}.}
	\label{fig:cation-concentration}
\end{figure}

Figure \ref{fig:cation-concentration} shows LSVs of the nanostructured electrode with a structure size of $\qty{75}{\nano\meter}$ in the mentioned electrolytes. Note that the current density is plotted in logarithmic scale and the non-faradaic current present at $\qty{0}{\RHE}$ has been subtracted from the data. It can be observed that highest activity is obtained in the most diluted electrolyte, i.e. with the lowest concentration of $\ce{K}$ cations in solution, while the lowest activity is found for the strongest electrolyte. The difference in overpotential between the least and the most dilutes electrolyte is approx. $\qty{35}{\milli\volt}$ at a HER current density of $\approx \qty{0.03}{\milli\ampere\per\square\centi\meter}$ (cf. orange arrow in the figure).

An opposite trend is found in the literature for continuous Au-layer electrodes. Here, at moderately alkaline pH, the highest HER activity is found for the strongest electrolyte \cite{Goyal2021}, i.e. electrolyte with highest concentration of cations. Thus the behavior of nanostructured electrodes is, again, opposite to the behavior known from continuous layer electrodes.

\subsection{Increased HER rate on various metal catalyst materials}

To further analyze HER rate enhancement on nanostructured electrodes, we consider two other  catalyst materials in addition to Au: Cu and Pt.
The Cu surface is produced by electrochemical copper deposition from solution onto the Au surface of the three electrode systems discussed above. Details of the deposition process can be found in appendix \ref{app:cu-deposition-stripping}.
The Pt nanostructres are fabricated analogously to the Au ones, i.e. by directly evaporating Pt instead of Au during the fabrication process of the electrodes. Typical CVs of the evaporated Pt surface in acidic as well as alkaline electrolyte demonstrating its quality are shown in appendix \ref{app:CVs-Pt-film}.

Figure \ref{fig:metal_comparison} shows LSVs of the three electrode systems with either (a) a Cu-based surface or (b) a Pt-based surface. Note that the HER activity of the Cu-based electrodes is measured in neutral electrolyte (pH 7, $\qty{0.18}{\molar}~\ce{NaOH}$ + $\qty{0.12}{\molar}~\ce{H2PO4^-}/\ce{HPO4^{2-}}$, Ar sat.'d),
while the HER activity of the Pt-based electrodes is measured in alkaline electrolyte (pH 13, $\qty{0.1}{\molar}~\ce{NaOH}$, Ar sat.'d).

\begin{figure}[h!]
	\centering
	\includegraphics[width=0.49\textwidth]{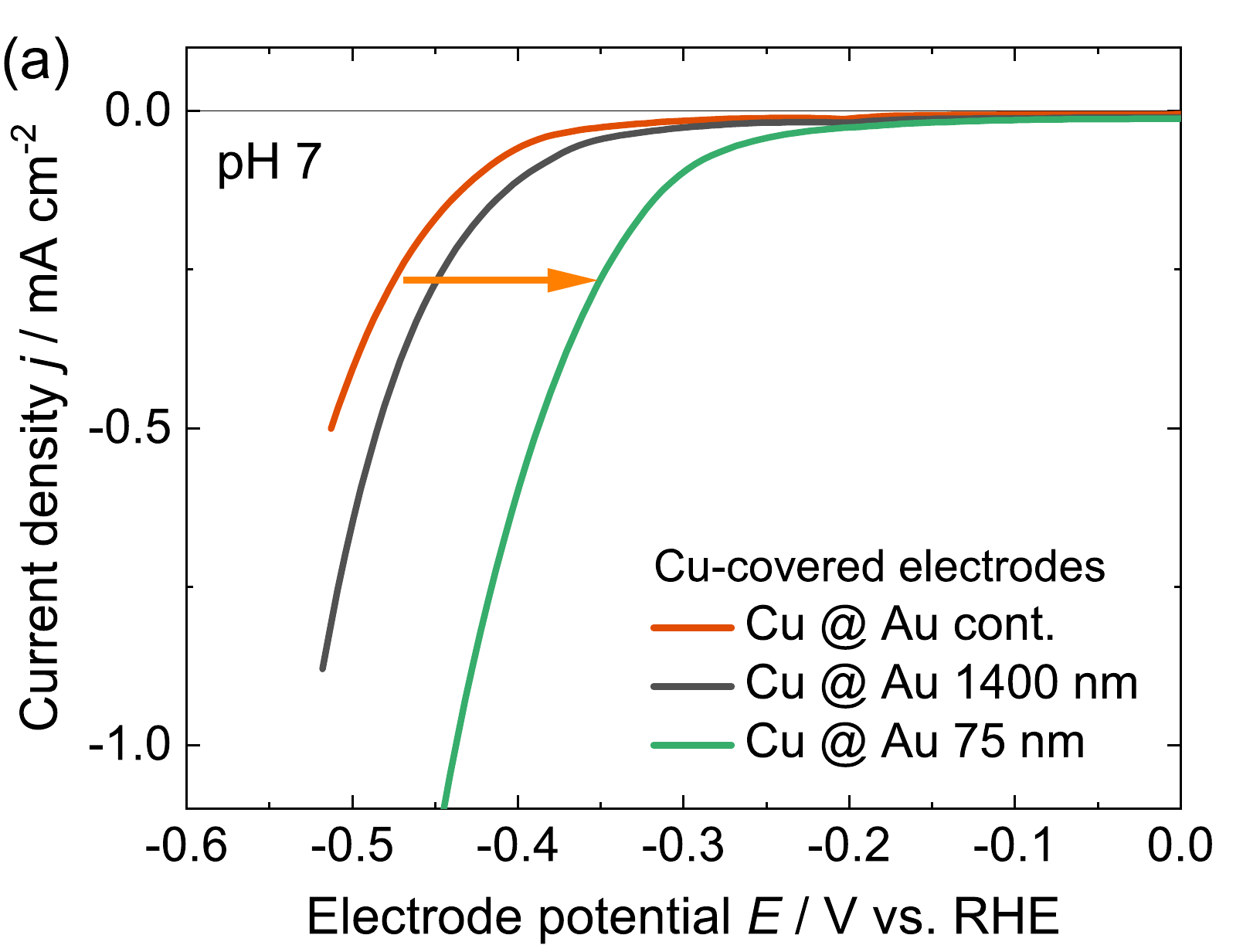}
	\includegraphics[width=0.49\textwidth]{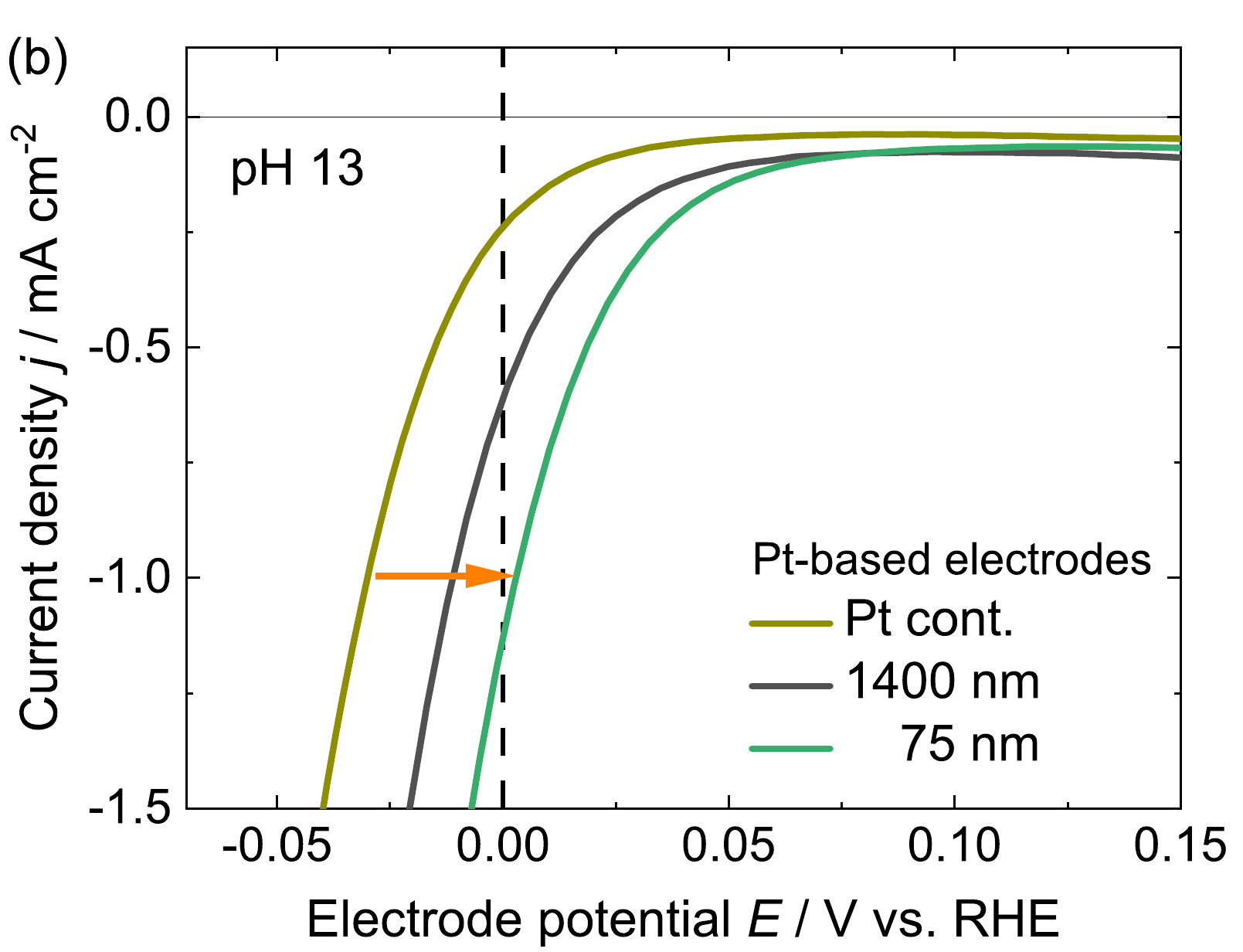}
	\caption{LSVs in the HER region for continouus metal layer electrodes and nanostructured electrodes. (a) Cu-covered electrodes in neutral electrolyte (pH 7, $\ce{formula}$/Ar sat.'d). (b) Pt-based electrodes in alkaline electrolyte (pH 13, $\qty{0.1}{\molar}~\ce{NaOH}$/ Ar sat.'d). Note that the scans are conducted in Ar saturated electrolyte, which gives the notable HER current already at positive potentials in RHE scale on the Pt-based electrodes. }
	\label{fig:metal_comparison}
\end{figure}

For both catalyst materials, we observe the same trend in HER activity for the three electrode systems. The continuous layer electrode system exhibits the lowest HER activity, while the system with the highest contribution of the metal/insulator interface (i.e. 75 nm structure size, green curves) exhibits the highest HER activity. This is indicated by the orange arrows in the Figures as a guide to the eye. This increase observed in Cu- and Pt-based systems is, thus, qualitatively similar to the increase found for Au-based electrodes, cf. Figure \ref{fig:size-series_si3n4}(a). However, quantitatively the increase is found to be different for the different metals.


\section{Discussion}\label{sec:discussion}



As shown in the result section, the HER activity of nanostructured electrodes increases when increasing the contribution of the metal/insulator interface (i.e. by decreasing the nanostructure size at similar coverage, cf. Figure \ref{fig:size-series_si3n4}). This suggests that the improved HER rate is due to an altered reaction mechanism at the metal/insulator interface, which was already suggested in one of our previous studies for gold/silicon oxide interfaces \cite{Maier2020}. Our results in this work show that the enhancement is present for different metal catalysts (Au, Cu, Pt) and that the degree of enhancement depends on several electrolyte properties, namely the type of cationic species, the cation (bulk) concentration and the pH of the electrolyte. These observations allow us to pinpoint the HER mechanism at the metal/insulator interface and thus the origin of the HER rate increase.


All data can be interpreted consistently, when assuming that the HER enhancement at the metal/insulator interface is caused by a combination of two factors: First, the hydrogen adsorption process is altered due to the presence of silanol (\ce{SiOH}) groups on the insulator surface. Second, charge transfer through the double layer is enhanced because the double layer above the insulator surface is less rigid. We will first explain these two parts of the mechanism and then discuss how it relates to the cation and pH trends observed experimentally.

\subsection{Altered hydrogen adsorption mechanism}
\begin{figure}[h!]
	\centering
	\includegraphics[width=0.7\textwidth]{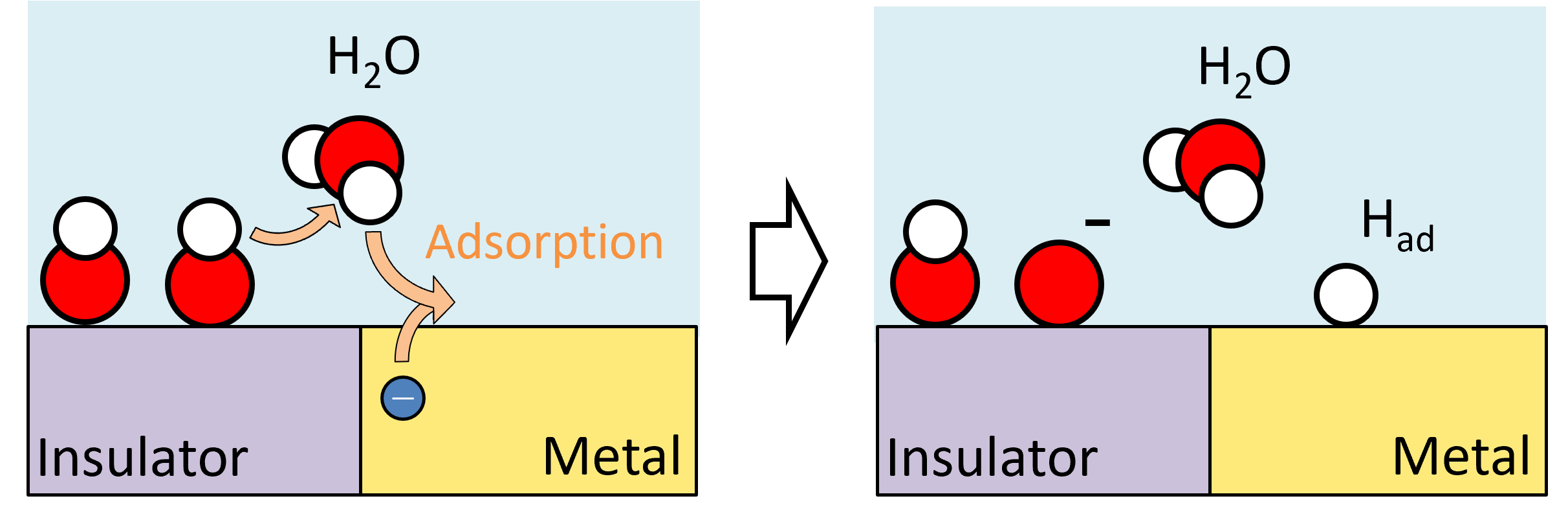}
	\caption{Scheme of the altered hydrogen adsorption process at the metal/insulator interface.}
	\label{fig:HER-bifunctional-mechanism_step1}
\end{figure}

In the Volmer step \eqref{eq:Volmer}, water molecules are dissociated into an (adsorbed) hydrogen atom and a hydroxide ion. At the metal/insulator interface, though, hydrogen atoms may be adsorbed by dissociating silanol groups (\ce{SiOH}) present on the insulator surface:
\begin{equation}
    \ce{SiOH + M^* + e^- -> SiO^- + M-H_{ad}}.
\end{equation}
A schematic of the process is shown in Figure \ref{fig:HER-bifunctional-mechanism_step1}.

Silanol groups are acidic, and in alkaline medium they dissociate according to the equilibrium
\begin{equation}
    \ce{SiOH + OH^- <=> SiO^- + H_2O}
\end{equation}
Reported $pK_\mathrm{a}$ values for this reaction on silica range between 6 and 11 \citep{van1996general, azam2012specific, backus2021probing}. These values are smaller than the $pK_\mathrm{a} = 14$ of water. This difference suggests that adsorbing a proton by dissociating silanol is easier than dissociating water. In addition, protons from silanol groups further away from the interface may be transported laterally across the surface through the Grotthuss mechanism (`bond flipping' of water molecules).

\begin{figure}[h!]
    \centering
    \includegraphics[width=\linewidth]{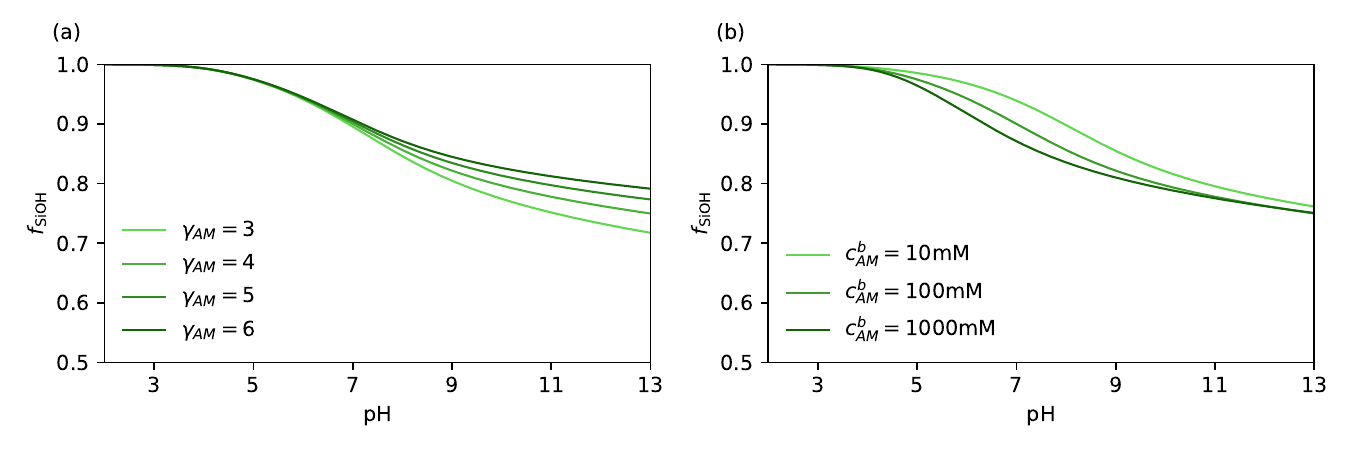}
    \caption{Fraction of protonated sites on the insulator surface, $f_\mathrm{SiOH}$, at different pH as obtained from our double layer model described in the methods section for (a) different effective cation sizes $\gamma_{AM}$ and (b) cation bulk concentrations $c_{AM}^\mathrm{b}$. Unless otherwise specified in the legend, $\gamma_{AM}=4$ and $c_{AM}^\mathrm{b}=100$ mM.}
    \label{fig:insulator-ph-sweep}
\end{figure}

The $pK_\mathrm{a}$ of silanol groups raises the question how many \ce{SiOH} groups are available, especially in an alkaline electrolyte. We computed the fraction of silanol sites that are protonated, $f_\mathrm{SiOH}$, using a double layer model from \citet{huang2021cation} and \citet{iglivc2019differential} (see Methods). The dependence of $f_\mathrm{SiOH}$ on pH is shown in Figure \ref{fig:insulator-ph-sweep} -- we discuss the dependence on effective cation size and concentration later. The important result is that over 70 percent of the silanol sites remain protonated, even at high pH. A fraction of between 70 and 80 percent protonated sites at alkaline pH was also reported by \citet{Zhang-pp154}. Clearly, $f_\mathrm{SiOH}$ is much higher than one would expect based on the $pK_\mathrm a$. The reason is likely the electrostatic repulsion of hydroxide ions from the negatively charged surface.

More arguments supporting the proposed altered adsorption process based on the experimentally observed cation and pH trends are given in section \ref{sec:discussion-cation-ph-trends}.


\subsection{Enhanced hydroxide ion transport through the double layer} \label{sec:discussion-mech2-transport}
The altered hydrogen adsorption process explains (at least in part) the increase in current for nanostructured electrodes with Au and Cu metal catalysts. However, Pt has a lower activation energy for hydrogen adsorption \cite{Schmickler2010}. Thus, an altered adsorption mechanism should not considerably affect the rate of Pt based electrodes. Instead, the removal of hydroxide ions from the Pt electrode surface appears to limit the HER current \cite{LedezmaYanez2017, bender2022understanding}. As mentioned in the introduction, \cite{LedezmaYanez2017} suggested that this is due to the difficulty of reorganizing the strongly polarized and concentrated electrolyte in the double layer. The fact that we do measure a larger current for laterally structured Pt/silicon nitride electrodes, cf. Figure \ref{fig:metal_comparison}(b), suggests that hydroxide ion transport is also enhanced on nanostructured electrodes.

\begin{figure}[h!]
	\centering
	\includegraphics[width=0.5\textwidth]{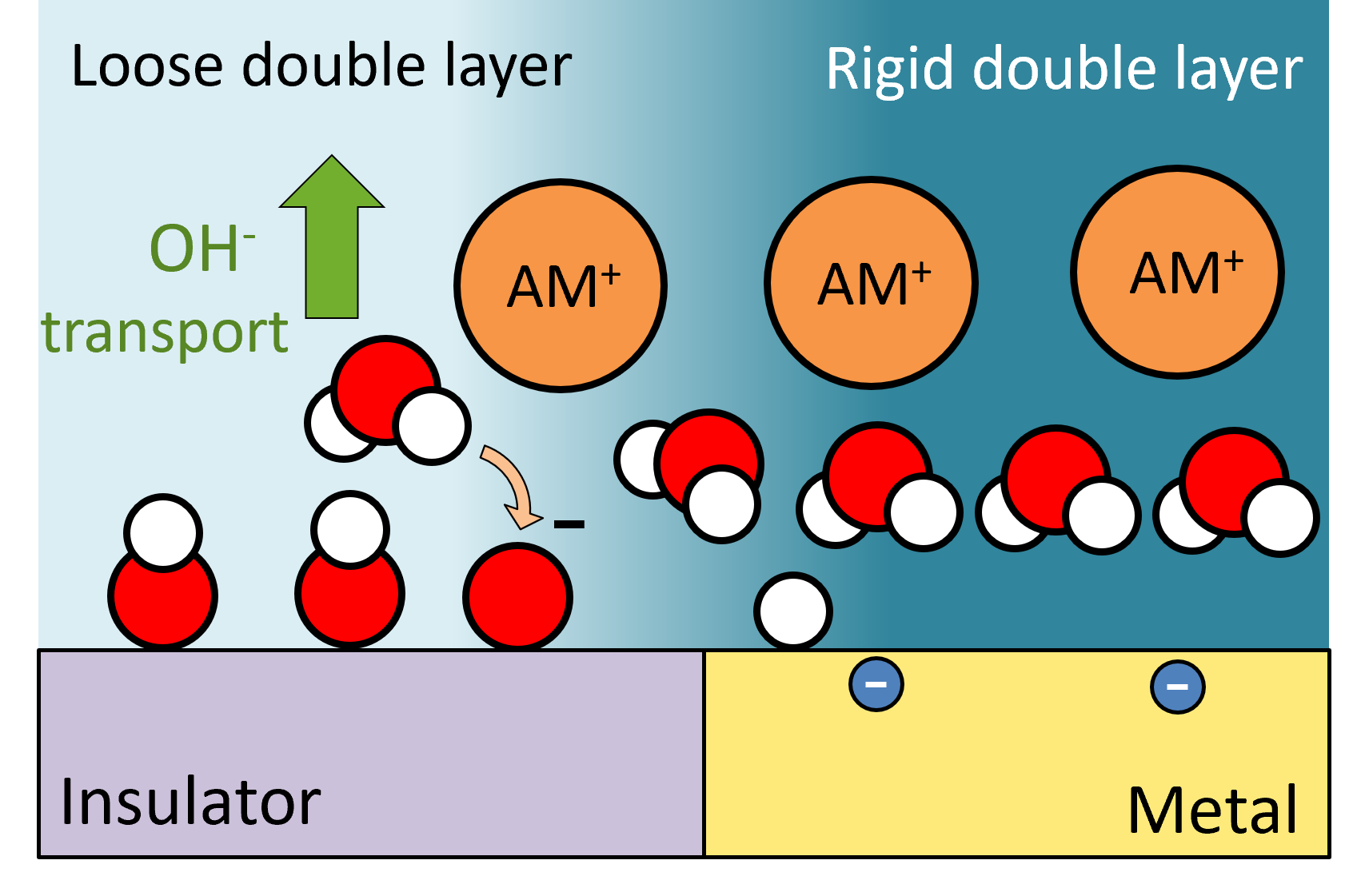}
	\caption{Scheme of the proposed enhanced \ce{OH-} transport process into the electrolyte at the insulator/electrolyte interface due to the lower rigidity of the double layer.}
	\label{fig:HER-bifunctional-mechanism_step2}
\end{figure}

We believe that this enhancement is due to a change in the spatial location of water dissociation. After a hydrogen atom from a silanol group is adsorbed, water molecules above the insulator surface are dissociated to protonate the silanol groups again and thereby complete the Volmer step:
\begin{equation}
    \ce{SiO- + H2O <=> SiOH + OH- (bulk)}
\end{equation}
The transfer of hydroxide ions from the electrode surface to the electrolyte bulk thus takes place above the insulator surface, rather than above the metal. At the metal/electrolyte interface, the entire applied voltage  drops across the double layer. At the insulator/electrolyte interface, on the other hand, most of the applied voltage drops across the insulator. The electric potential at the insulator surface then depends mostly on the charge from \ce{SiO-} groups. The potential at the insulator is in general much lower than at the metal. Consequently, we expect that the electrolyte at the insulator/electrolyte interface is less polarized and concentrated, i.e. less `rigid'. This lower rigidity, in turn, should facilitate water dissociation and improve the transport of $\ce{OH-}$ to the electrolyte bulk. This effect is illustrated schematically in Figure \ref{fig:HER-bifunctional-mechanism_step2}.

We quantify the notion of double layer rigidity using the pressure due to electrostatic forces in the double layer as previously considered in refs. \cite{dreyer2013overcoming, dreyer2014mixture,landstorfer2016theory, landstorfer2022thermodynamic}. For a relative permittivity $\varepsilon$ that depends on the concentration of dipoles and the electric field $\mathcal E$,  the pressure gradient at distance $x$ from the electrode is given by \cite{landstorfer2022thermodynamic}
\begin{equation} \label{eq:gradp}
    \frac{\partial P}{\partial x} = \rho_\mathrm{F} \mathcal E + \varepsilon_0 (\varepsilon - 1) \mathcal {E} \frac{\partial \mathcal E}{\partial x}
\end{equation}
where $\rho_\mathrm F$ is the free charge density due to ions in the electrolyte. Equation \ref{eq:gradp} can be integrated to yield the pressure $P$ at the reaction plane (see Methods). We again use the double layer model to calculate $\mathcal E$, $\rho_\mathrm{F}$ and $\varepsilon$ and consequently the pressure $P$ above a metal (here Au) surface and above an insulator surface. The effective sizes of cations (including solvation shells) are specified using size factors $\gamma_{AM}$ -- larger $\gamma_{AM}$ means a larger degree of solvation. For the insulator, we assume that the applied potential $\phi_0$ drops completely over the insulator layer, meaning that the double layer structure is not affected by changes in the applied potential. This is a reasonable assumption as the actual thickness of the insulator layer is much larger than the thickness of the corresponding Helmholtz layer.

\begin{figure}[t]
	\centering
	\includegraphics[width=\linewidth]{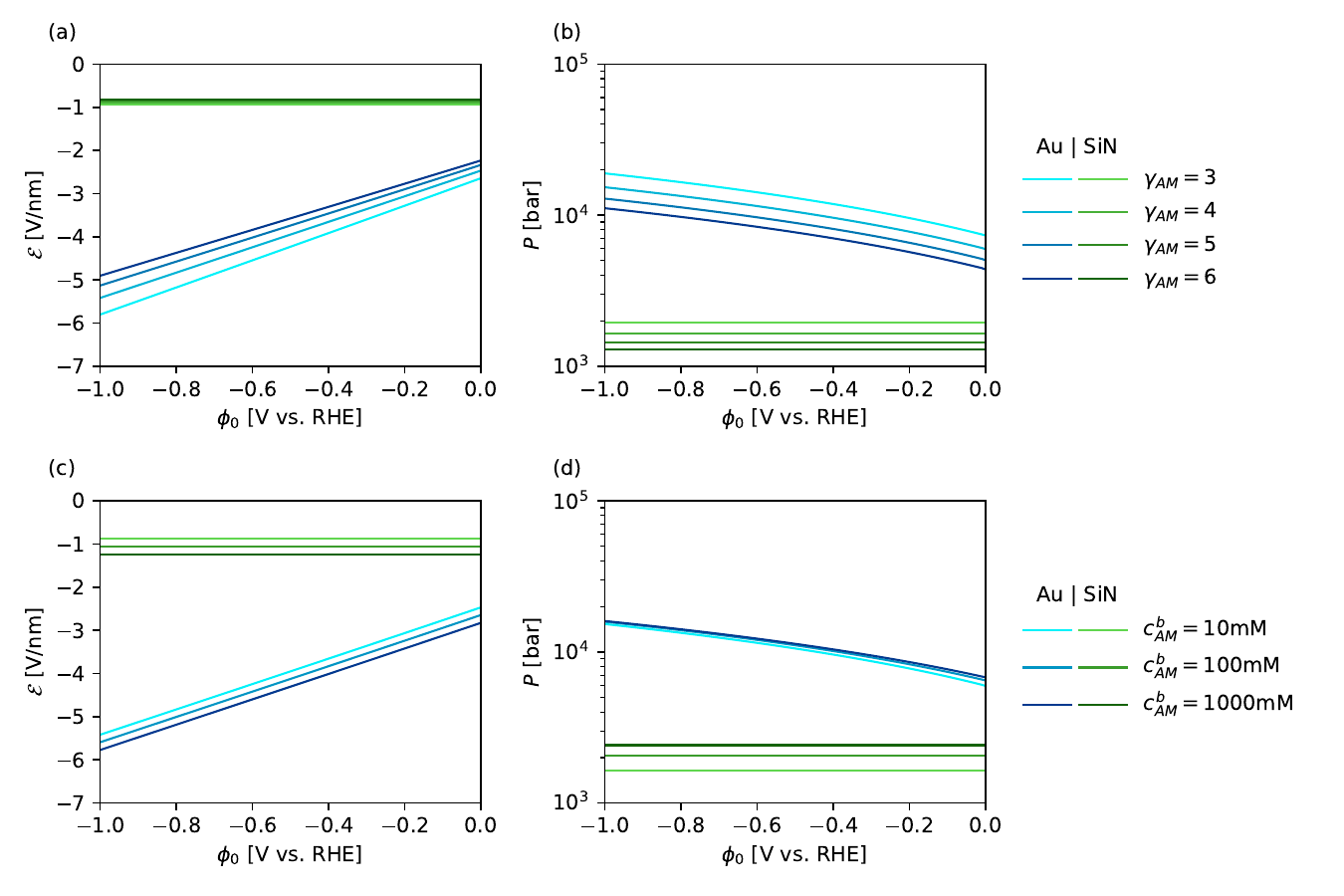}
	\caption{Electric field $\mathcal{E}$ and pressure $P$ in the electrolyte near silicon nitride insulator and metal surfaces. The curves for the insulator are indicated in green, and the curves for the gold are blue. (a), (b): electric field and pressure plotted against the potential applied at the electrode, $\phi_0$, for different effective cation sizes $\gamma_{AM}$. (c), (d): electric field and pressure plotted against the applied potential for different cation bulk concentrations $c_{AM}^\mathrm{b}$. Unless otherwise specified in the legend, $\gamma_{AM}=4$ and $c_{AM}^\mathrm{b}=\qty{10}{mM}$. The pH is 13.}
	\label{fig:pressures}
\end{figure}

The electric field behavior plotted over the applied potential is shown in Figure \ref{fig:pressures} for (a) various cation sizes $\gamma_{AM}$ and (c) various bulk concentrations of cations $c_{AM}^\mathrm{b}$. The respective results of the calculated pressure are shown in Figures \ref{fig:pressures} (b) and (d).
Comparing the pressure at the Au surface (blue lines) to the pressure at the insulator surface (green lines), we can immediately see that the pressure at Au is an order of magnitude higher. The pressure force that hydroxide ions need to overcome to be transported to the electrolyte bulk is thus indeed smaller when water is dissociated at the insulator surface rather than at the metal surface.
This result explains the enhancement on Pt nanostructured electrodes. It also explains why, despite the enhancement of proton adsorption, Au nanostructured electrodes do not become limited by ion transport. We know that the adsorption step is still rate-limiting because the Tafel slope for nanostructured Au electrodes is the same as for continuous gold electrodes, cf. Figures \ref{fig:size-series_si3n4}(b) and \ref{fig:cation-species}(b).

\subsection{Experimental cation and pH trends}\label{sec:discussion-cation-ph-trends}
Next, we check if the proposed mechanism is in accordance with the experimentally observed cation and pH trends. According to Koper and co-workers \citep{Goyal2021, Monteiro2021}, cations in the double layer reduce the activation barrier of water dissociation. However, the dissociation of silanol groups rather than water molecules is a different mechanism. Consequently, the added benefit of cations may be lost, and we can expect the cation trends to change, which we indeed observe: At the Au/insulator interface, a higher HER activity is obtained for more strongly hydrated cation species (Figure \ref{fig:cation-species}) and for a lower cation bulk concentration (Figure \ref{fig:cation-concentration}), and vice versa. Also, the nanostructured electrodes showed a greater enhancement at pH 7 than at pH 13 (Figure \ref{fig:pH-value}). All those trends can be consistently explained when assuming that the HER current at the metal / insulator interface is determined by the concentration of the silanol groups and the rigidity of the double layer.


\paragraph{Availability of protonated silanol.} Silanol groups (\ce{SiOH}) are a reactant in the proposed mechanism, and so the measured current should scale with $f_\mathrm{SiOH} = \bar n_\mathrm{SiOH} / (\bar n_\mathrm{SiO^-} + \bar n_\mathrm{SiOH})$, with $\bar n$ denoting a surface number density. As shown in the simulation results in Figure \ref{fig:insulator-ph-sweep}, this fraction depends on the pH, the cation size, and the cation concentration. We conclude that more strongly hydrated cations (large effective size) and smaller bulk cation concentrations yield the highest $f_\mathrm{SiOH}$. The current we measured experimentally is indeed higher under circumstances where $f_\mathrm{SiOH}$ is also higher. In addition, we can now explain the pH trend we measured: At a lower pH, the concentration of \ce{SiOH} is higher, resulting in a higher current density.

\paragraph{Double layer rigidity.} In the pressure calculations shown in Figure \ref{fig:pressures}, we see that a lower cation bulk concentration $c_{AM}^\mathrm{b}$ yields lower pressures above the insulator. In addition, the pressure is lower for larger effective cation size $\gamma_{AM}$. Weakly solvated cations simply pack more tightly, allowing for a larger cation concentration at the surface, which increases the pressure according to equation \eqref{eq:gradp}. A higher pH increases the surface charge and consequently the electric field (see Figure \ref{fig:insulator-ph-sweep}), which also increases the pressure above the insulator. Altogether, the current we measured experimentally correlates with lower pressures in the double layer above the insulator.

\paragraph{Lateral proton transfer.} An additional aspect that should be considered is the effect of the double layer structure on the transport of protons from silanol groups to the metal catalyst. Proton transport through the Grotthuss mechanism requires that water molecules can orient themselves to form a hydrogen bonding network, which may be easier in a weaker electric field. Strongly solvated cations would disturb this water network less. We again find the same trend: more strongly solvated cations, lower cation bulk concentrations, and a weaker electric field improve the lateral transport of protons and therefore their availability at the metal surface. Although the effect is different from the transport of hydroxide ions, it appears likely that the relevant quantities are combined in the pressure, and influence the current in the same way. Further investigations will show whether this assumption is correct.


\subsection{Qualitative model of HER current based on electrostatic pressure}
To further strengthen our hypothesis we present a qualitative model of the HER current based on the electrostatic pressure in the following. The model is used to elaborate the theoretical dependence of the HER current on the relevant experimental parameters. Although a full theoretical description of the proposed mechanism is out of the scope of this work, we propose a first qualitative description.

According to Marcus theory, the activation energy $\Delta G_\ddag$ of an electrochemical reduction reaction depends on the overpotential $\eta$ and the reorganization energy $\lambda$ as
\begin{equation} \label{eq:marcus}
    \Delta G_\ddag = \frac{(\lambda + e \eta)^2}{4\lambda}
\end{equation}
\cite{Schmickler2010, huang2020mixed}, here $e$ is the elementary charge. The overpotential is defined as $\eta = \phi_0 - E_\mathrm{eq}$ where $E_\mathrm{eq}=\qty{0}{V}$ is the equilibrium potential of HER given by the Nernst equation and $\phi_0$ is the electrode potential (in V vs. RHE). The reorganization energy is typically on the order of electron volts \cite{Schmickler2010}.

In standard Marcus theory, $\lambda$ is usually considered to be the reorganization of the dielectric due to the change in charge on the reactant \cite{Schmickler2010, huang2020mixed}. Here, in the context of HER, we replace $\lambda$ by a more general `reorganization' energy $\Lambda$, which also takes into account any steric forces encountered by ions being transferred through the double layer:
\begin{equation}
    \Delta G_\ddag = \frac{(\Lambda + e \eta)^2}{4\Lambda}.
\end{equation}
Note that a linear Ansatz for $\Delta G_\ddag$ would give similar results, but we choose the form of equation \eqref{eq:marcus} to interpret new concepts in terms of the existing theory. As a simple model, we now consider a linear dependence of the reorganization energy on the pressure:
\begin{equation}
    \Lambda = C_1 e + C_2 (d_\mathrm{H_2O})^3  P
\end{equation}
where $(d_\mathrm{H_2O})^3$ is roughly the volume of a water molecule. $C_1$ has the unit Volt and $C_1 e$ can be considered to be the reorganization energy in the absence of steric forces. $C_2$ is unitless, and can be seen as representing the number of water molecules that have to be reorganized. Here, we choose $C_1=\qty{6}{V}$ such that $\Lambda > |e\eta|$ with some margin over the range of the studied parameters, and $C_2=3$ so that $C_2(d_\mathrm{H_2O})^3$ is roughly the volume of a solvated hydroxide ion.

Now, we use $P$ and $f_\mathrm{SiOH}$ calculated from the mean-field electrolyte model, and simulate the cathodic current density with an Arrhenius-type ansatz:
\begin{equation}
    j = -A f_\mathrm{SiOH} e^{-\beta \Delta G_\ddag}
\end{equation}
with $A$ a constant factor and $\beta = 1/(k_\mathrm{B} T)$ the inverse temperature. Because electron transfer and double layer reorganization are assumed to be spatially separated, we use the potential at the metal surface to calculate the overpotential $\eta$, but we use the pressure at the insulator surface to compute $\Lambda$ and consequently $\Delta G_\ddag$.

\begin{figure}[h!]
    \centering
    \includegraphics[width=\linewidth]{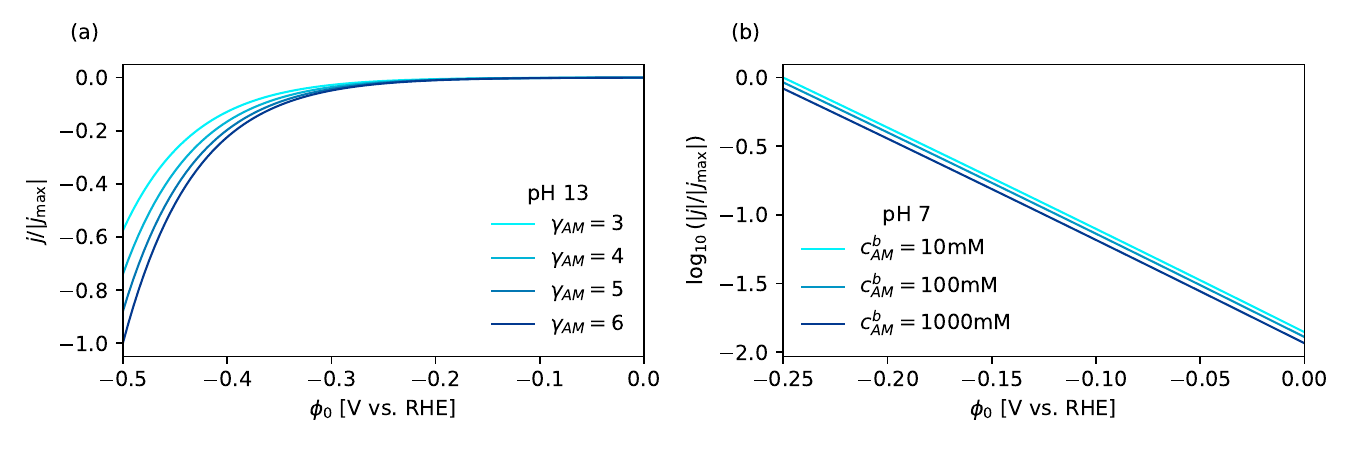}
    \caption{Current density $j$ plotted against applied potential $\phi_0$ for (a) different cation sizes $\gamma_{AM}$ and (b) different bulk concentrations $c_{AM}^\mathrm{b}$. The current density is always normalized to the largest absolute value in the plotted potential range, $|j_\mathrm{max}|$. Conditions are the same as for the experimental data: For (a), the pH is 13 and $c_{AM}^\mathrm{b}=\qty{0.1}{M}$ (cf. Fig. \ref{fig:cation-species}). For (b), we chose $\gamma_{AM}=4$; the pH is 7, cf. Fig. \ref{fig:cation-concentration}.}
    \label{fig:sim-cat-trends-result}
\end{figure}

Figure \ref{fig:sim-cat-trends-result} gives the dependence of the simulated current density $j$ as a function of the applied electrode potential (a) for various cation sizes $\gamma_{AM}$ and (b) for different bulk concentration of cations $c_{AM}^\mathrm{b}$. The cation concentration trend observed experimentally for nanostructured electrodes, cf. Figure \ref{fig:cation-concentration}, is qualitatively reproduced. For different ion sizes, the current increases in the order of scaling factors $F_{AM}$, cf. equation \eqref{eq:scaling-factor}.We attribute the difference from the actual current trends shown in Figure \ref{fig:cation-species} to the fact that the measured current is a superposition of current from the continuous gold surface, where HER occurs through water dissociation, and current from the gold/insulator interface regions, where HER occurs through silanol dissociation.

These theoretical predicted dependencies agree well
with the trends observed in the experimental data, supporting our proposed modified HER mechanism at the metal/insulator interface.

\subsection{Comparison to bifunctional HER mechanisms proposed by other groups}
Next, we compare the mechanism proposed here with bifunctional HER mechanisms proposed by other groups: Markovic et al. decorated metal surfaces, which fulfill the functionality of the hydrogen adsorption site ('catalyst'), by metal hydroxide particles ('support'), such as $\ce{Ni(OH)2}$ \cite{Subbaraman2011, Danilovic2012} or other 3d transition metals \cite{Subbaraman2012, Strmcnik2016}. The authors elucidated a reaction mechanism in which the transition metal(hydroxide) serves to enhance the water dissociation step by adsorbing the $\ce{OH}_\mathrm{ad}$ intermediate on the transition-metal while the $\ce{H}_\mathrm{ad}$ intermediate adsorbs on the metal catalyst surface.
\citet{McCrum2020} further elucidated the role of the supporting metal, since they found a volcano shaped curve for the HER activity on the bifunctional interface as a function of the hydroxide adsorption energy of the support metal, which makes the $\ce{OH}_\mathrm{ad}$ binding energy an additional descriptor.

However, we do not believe that this bifunctional mechanism is present in our system. The oxophilicity of the silicon-based support material we use here is rather strong \cite{Kepp2016}. According to Markovic et al. \cite{Subbaraman2012, Strmcnik2016} a too-strong binding of the $\ce{OH}_\mathrm{ad}$ intermediate would lead to a 'poisoning' of the support surface. Consequently, the desorption of the $\ce{OH-}$ is hindered and the support would further behave as a spectator species in the reaction.
In addition, if the silicon-based material would behave as a hydroxide adsorption site in the reaction, the observed dependence of the HER activity on the cationic species would be different. It is known that the presence of the alkali-metal cation weakens the hydroxide adsorption \cite{Chen2017}. The hydroxide adsorption is weakened least for Li and most for Cs \cite{Chen2017}. Since a weakening of the hydroxide adsorption would be beneficial for a strongly adsorbing material such as Si, we would expect the highest improvement when using Cs as the alkali metal cation. However, we see a reverse effect, as Li gives the highest improvement for the metal/insulator interface.

We believe that the HER enhancement observed in our system comes from the local decoupling of the hydrogen adsorption on the metal surface, and the water dissociation at the insulator (i.e. 'support') interface. The transferred species between catalyst and support is likely a proton.

\section{Conclusion}\label{sec:conclusion}
In conclusion, we present here a novel mechanism, which enhances the Hydrogen Evolution Reaction from water reduction at a metal/insulator interface. We observed this mechanism by investigating electrodes consisting of gold nanostructure arrays embedded in an insulating layer (here silicon nitride) on a silicon-based substrate. By changing the diameter of the individual structures in the array, we saw that the border of the structures (i.e. the metal/insulator interface) exhibits a much larger exchange current density than the metal bulk part of the structures. This enhancement is not just present for gold/silicon nitride interfaces, but also occurs in copper/silicon nitride and platinum/silicon nitride systems, showing the generality of the mechanism.

The dependence of the HER activity on electrolyte properties is altered at the metal/insulator interface compared to the dependence known for bare metal surfaces:
At the metal/insulator (here gold/silicon nitride) interface, the HER activity is larger if the pH is lower, the bulk cation concentration is lower, or the cation species changes to a species with a higher solvation energy. The latter is observable by comparing the scaling factors $F$ of the exchange current densities for the various cationic species. We found  $F_\mathrm{Cs}\approx 12$ as the lowest and $F_\mathrm{Li}\approx 40$ as the largest scaling factor, i.e. improvement.

Based on these findings we elaborate a HER mechanism occurring at the metal/insulator interface: The silicon nitride (and in principle any insulator providing OH groups) can work as a proton source for HER occurring at the adjacent metal surface, which lowers the activation energy for adsorption, as no water dissociation is necessary in the double layer above the metal surface. The electrostatic pressure above the insulator layer is much lower compared to the metal surface and almost independent of the applied electrode potential, as most of the applied voltage drops across the insulator layer. Thus, water dissociation and the subsequent transport of $\ce{OH-}$ into the electrolyte bulk is facilitated above the insulator layer. Altogether, the two spatially separated reaction steps of the Volmer step can be written as:
\begin{alignat}{2}
	\ce{& SiOH + M^* + e- &&-> SiO- + M-H_{ad} } \label{eq:bifunctional-1}\\
	\ce{& SiO- + H2O &&-> SiOH + OH-}	\label{eq:bifunctional-2}
\end{alignat}
The appearance of this doubly bifunctional mechanism can strongly enhance the HER kinetics, since the reaction speed is only determined by the proton adsorption process at the metal surface. Any other processes originating from double layer properties, which may indirectly hinder the reaction, such as the transport of the $\ce{OH-}$ species through the rigid double layer into the electrolyte bulk \cite{LedezmaYanez2017} or the need for the dissociation of the water molecule in the double layer \cite{Goyal2021} above the metal surface, can be avoided.

The knowledge of this mechanism enables novel design possibilities for efficient electrocatalysts. The usage of insulators as a proton source for hydrogenation reactions is an easy method to solve drawbacks stemming from fundamental properties of electrochemical interfaces and it gives a cheap possibility to enhance the efficiency of metal-based catalysts. Equally important is the possibility to adjust the pressure, i.e. the rigidity of the double layer, above the insulator independently of the pressure above the metal, which opens a further route to optimize electrochemical reaction rates.


\section*{Methods}\label{sec:experimental}

\subsection*{Electrode fabrication}
The metal arrays on the electrodes investigated in this work are produced by Lift-off Nanoimprint Lithography (LO-NIL) \cite{Nagel2017, Maier2020, Golibrzuch2022}. The surface area covered by the nanostructure array expands over $5\times5 ~\unit{\centi\meter^2}$ for all nanostructured electrodes investigated here. Geometric properties of the arrays with various structure sizes can be found in the appendix \ref{app:geometric-properties}. The fabrication method itself is described in detail in some of our previous publications \cite{Maier2020}. The silicon substrates used for nanostructuring are prepared from a commercial silicon wafer (n-doped, FZ:(111), $1-\qty{10}{\ohm\centi\meter}$, Si-Mat Silicon Materials, Germany), which came already covered by a LPCVD silicon nitride ($\ce{Si3N4}$) layer with a thickness of $\qty{17}{\nano\meter}$. This layer is isotropically etched in an reactive ion etching process to the desired thickness of $\qty{12}{\nano\meter}$.

\subsection*{Electrochemical measurements}

The electrochemical cell used for the experiments is a custom-built, air tight three-compartment cell made of PCTFE. Its front is covered by a glass cover window made from hardened mineral glass. The three compartments (resp. containing working, reference and counter electrode) are separated from each other by a proton conducting membrane (Nafion$^\text{TM}$, Chemours, USA) to avoid cross-contamination between the compartments. The reference electrode used is a commercial Mercury/Mercurous Sulface (MSE) reference electrode (International Chemistry Co. Ltd., Japan) which contains saturated $\ce{K2SO4}$ solution. The measured electrode potential in MSE scale is converted in SHE scale using an offset of $\qty{0.64}{\volt}$ \cite{BardFaulkner2001}. We use a coiled gold-wire as the counter electrode.

All electrolytes used here are mixed from suprapur grade salts and  $\qty{18.2}{\mega\ohm\centi\meter}$ DI-Water (Elga Purelab, Veolia Water Technologies, Germany). The electrolytes are saturated with Ar gas (purity 5.0, Westfahlen, Germany) prior the electrochemical experiments.

Linear sweep voltammetric scans and cyclic voltammetric scans are usually conducted with the standard scan rate of $\qty{50}{\milli\volt\per\second}$. The Tafel analysis is performed by an analysis of the logarithmically plottet data of the scans. The fits were obtained in a potential window between $\qty{-0.25}{\RHE}$ and $\qty{-0.4}{\RHE}$. Tafel slopes and exchange current densities given here are averages over three individual measurements (i.e. electrodes).

\subsection*{Surface determination}
All measured currents are normalized to the active surface area of the respective electrode. The electrochemical active surface area of the Au-based electrodes was determined via the $\ce{OH}$ desorption in $\SI{0.1}{\molar}$ \ce{H2SO4} (suprapur grade, Merck) in cyclic voltammetric experiments. As an upper potential limit $\SI{1.75}{\RHE}$ was chosen, as it has been found to be the approximate potential at which a 'monolayer-like coverage' is achieved \cite{Hamelin1996}. The transferred charge during the subsequent gold oxide reduction process was divided by $\qty{390}{\micro\coulomb\per\square\centi\meter}$, which is the well accepted specific surface charge value for Au surfaces \cite{Trasatti1991, BardFaulkner2001}. In the appendix \ref{app:auox-reduction_utp}, exemplary CVs of a continuous Au-layer electrode in $\SI{0.1}{\molar} ~\ce{H2SO4}$ in the Au oxidation region and the determined charge transferred during the AuOx reduction process as a function of the upper turning potential is shown.

The surface of the Pt-based electrodes is determined via under potential H-ad- and desorption in a cyclic voltammetric experiment conducted in $\qty{0.1}{\molar}~\ce{H2SO4}$/Ar sat.'d. We chose $\qty{0.04}{\RHE}$ as the lower potential limit and $\qty{0.5}{\RHE}$ as the upper potential limit of the scan conducted with $\qty{50}{\milli\volt\per\second}$. The average transferred charge of H-UPD and H-stripping is divided by the specific Pt surface charge of $\qty{210}{\micro\coulomb\per\square\centi\meter}$ \cite{Trasatti1991} to calculate the active surface area.

We consider the active surface area of the Cu-covered electrodes to be similar to the active surface area of the corresponding underlying Au-based electrode, as we assume that the electrochemical coverage of the Au-surface does not alter the respective active surface area tremendously.

The surface determination of the individual electrodes has been carried out before and after the experimental procedures performed. It turned out that the change during the experimental procedure never exceeded 5\%.

\subsection*{Electrolyte model} \label{sec:methods-model}
We model the local reaction conditions in the electric double layer using theory from \citet{iglivc2019differential} and \citet{huang2021cation}. In this theory, the electric potential $\phi$ is described by a Poisson-Boltzmann equation, which reads
\begin{equation} \label{eq:mpbe}
    -\frac{\partial}{\partial x} \left[\varepsilon_0 \varepsilon(x)  \frac{\partial \phi}{\partial x} \right]
    =\sum_i z_i e n_i(x).
\end{equation}
Here, $\varepsilon_0$ is the vacuum permittivity, $\varepsilon$ the relative permittivity, $e$ the elementary charge, and $z_i$ and $n_i$ are the charge number and number density of species $i$ in the electrolyte, respectively. We include protons, hydroxide ions, alkali metal cations $AM$, anions $X$, and water molecules: $i \in \{\mathrm{H^+}, ~\mathrm{OH^-}, ~AM, ~X, ~\mathrm{H_2O}\}$. An overview of the parameters in the model can be found in Table \ref{tab:parameter-table}.

The relative permittivity depends on the local electric field $\mathcal E=-\partial \phi/\partial x$ and number density of (polarizable) water molecules $n_\mathrm{H_2O}$ as
\begin{equation} \label{eq:permittivity}
    \varepsilon (x) = \varepsilon_\infty + \frac{p n_\mathrm{H_2O}(x) }{\mathcal{E}(x)} \mathcal{L}(\beta p \mathcal{E}(x))
\end{equation}
with $\varepsilon_\infty$ the optical permittivity, $p$ the dipole moment, and $\mathcal L(y) = \coth (y) - 1/y$ the Langevin function. To find $p$ consistent with the bulk permittivity of water $\varepsilon_\mathrm{w}\approx 78.5$, we note that $\mathcal E=0$ in the bulk. From equation \eqref{eq:permittivity} it then follows that $\varepsilon_\mathrm{w} = \varepsilon_\infty + \frac13 \beta p^2 n_\mathrm{w}$, or
\begin{equation}
    p = \sqrt{\frac{3(\varepsilon_\mathrm{w} - \varepsilon_\infty)}{\beta n_\mathrm{w}}}
\end{equation}
with the bulk number density of pure water $n_\mathrm{w} = \SI{55.5}{M} \times N_\mathrm{A}$.

The finite size of ions in the electrolyte is accounted for by considering a lattice with lattice spacing $d_\mathrm{H_2O}$, the effective diameter of a water molecule. This effective diameter is calculated such that $n_\mathrm{w}=(d_\mathrm{H_2O})^{-3}$ -- this yields $d_\mathrm{H_2O}\approx3.1$ Å. Ions may occupy multiple lattice sites; their effective (i.e. including solvation shell) sizes are specified using the relative size factor
\begin{equation}
    \gamma_i = \left(\frac{d_i}{d_\mathrm{H_2O}}\right)^3
\end{equation}
with $d_i$ the effective diameter of species $i$. Ions with a high degree of solvation such as \ce{Li+} have a large effective size; weakly solvated ions like \ce{Cs+} have a small effective size. \citet{drab2017internal} inferred from bulk permittivity data that $\gamma_{AM} + \gamma_{X} \approx 7.5$ for \ce{NaCl} electrolytes. Because the solvation shells of ions are not fixed, it is difficult to connect values of $\gamma$ to the actual size of ions. Here, we investigate cation size trends qualitatively by considering $\gamma_{AM} = 3, 4, 5, 6$, which are considered to be in the range of the cations used experimentally.

Using these size factors, the number densities are written as a fraction of the lattice site density $n_\mathrm{max}=(d_\mathrm{H_2O})^{-3}$ ($=n_\mathrm{w}$) as
\begin{equation} \label{eq:model-number-densities}
    n_i = n_\mathrm{max} \frac{\chi_i \Theta_i(x)}{\sum_j \gamma_j \chi_j \Theta_j(x)}
\end{equation}
where $n_i^\mathrm{b}$ is the bulk number density of species $i$ and $\chi_i = n_i^\mathrm{b}/n_\mathrm{max}$. $\Theta_i$ are Boltzmann factors, defined as
\begin{equation}
    \Theta_i(x) = \begin{dcases}
        \exp(-\beta z_i e \phi(x)), &\text{for $i \in \{\mathrm{H^+, OH^-,} AM, X \}$;}\\
        \frac{\sinh{\beta p \mathcal{E}(x)}}{\beta p \mathcal{E}(x)}, &\text{for $i=\mathrm{H_2O}$.}
    \end{dcases}
\end{equation}
Note that if $\Theta_i \gg \Theta_j$ for all $j$, $n_i$ approaches a saturation limit concentration of $n_\mathrm{max}/\gamma_i$.

As potential reference, we choose the point of zero charge (PZC) of gold, which was measured to be $E_\mathrm{PZC,Au}=\qty{0.2}{\SHE}$ for the electrodes considered in this work \citep{Maier2023}. A potential with respect to this PZC can be converted to a potential with respect to RHE as
\begin{equation}
    \phi \text{(vs. RHE)} = \phi \text{(vs. PZC)} + E_\mathrm{PZC,Au} + \qty{59}{mV} \times \mathrm{pH}.
\end{equation}

The boundary conditions to equation \eqref{eq:mpbe} are specified as follows. First, in the solution bulk, $\phi = 0$ V vs. PZC. At the electrode, the boundary conditions are specified on the plane of closest approach for ions at distance $x_2$ from the electrode surface. Here, we consider $x_2= d_{AM}/2$, as we only consider potentials negative of the point of zero charge of gold and so the plane of closest approach is set by the size of the cations. The potential at $x_2$ is denoted $\phi_2:=\phi(x_2)$.

\paragraph{Metal boundary condition.} Between the electrode surface and $x_2$, there are no free charges, so the potential profile is linear. For metal surfaces at potential $\phi_0$, the boundary condition then reads
\begin{equation} \label{eq:metalbc}
    -\mathcal E(x_2)x_2 = \phi_2 - \phi_0.
\end{equation}

\paragraph{Insulator boundary condition.} \citet{van1996general} derived the surface charge at insulator surfaces with acidic surface OH groups as
\begin{equation}
    \sigma = - e \bar{n}_\mathrm{sil} \frac{K_\mathrm{a}}{K_\mathrm{a} + c_\mathrm{H^+}^\mathrm{b} \exp(-\beta e \phi_0)}.
\end{equation}
Here, $c_\mathrm{H^+}^\mathrm{b}$ is the bulk concentration of protons (although it is small in alkaline medium, it gives the same result as when rewriting the fraction in terms of $c_\mathrm{OH^-}^\mathrm{b}$). $\bar{n}_\mathrm{sil}= 5\times 10^{18} \qty{}{m^{-2}}$ is the areal density of silanol (\ce{SiOH} and \ce{SiO^-}) sites and $K_\mathrm{a}=10^{-6}$ M is the acid dissociation constant for silanol -- these values are used for silica by \citet{van1996general}; silicon nitride should behave similarly. Together with equation \eqref{eq:metalbc} and $\sigma =  \varepsilon_0 \varepsilon(0) \mathcal{E}(0) = \varepsilon_0 \varepsilon(x_2) \mathcal{E}(x_2)$, the boundary condition for silica surfaces is finally
\begin{equation}
    \varepsilon_0 \varepsilon(x_2) \mathcal{E} (x_2) =  - e \bar{n}_\mathrm{sil} \frac{K_\mathrm{a}}{K_\mathrm{a} + c_\mathrm{H^+}^\mathrm{b} \exp(-\beta e (\phi_2 + \mathcal{E}(x_2)x_2))}.
\end{equation}
The fraction $f_\mathrm{SiOH}$ of protonated silanol sites is calculated as $f_\mathrm{SiOH} = (1 - \frac{-\sigma/e}{\bar n_\mathrm{sil}})$.

Details on the numerical implementation of the model can be found in Appendix \ref{ssec:model-numerical-solution}.

\subsection*{Calculation of the pressure}
As mentioned in the main text, \citet{landstorfer2022thermodynamic} derived that for a concentration- and electric field-dependent permittivity, the pressure gradient at $x$ is given by equation \eqref{eq:gradp}. The free charge density due to electrolyte ions is $\rho_\mathrm F = \sum_i n_i z_i e$. Equation \eqref{eq:gradp} can be integrated to yield the pressure at the reaction plane,
\begin{equation}
    P(x_\mathrm{rp}) = P_0 -\int_{x_2}^\infty \frac{\partial P}{\partial x} \mathrm d x
\end{equation}
where $P_0$ is the pressure in the bulk of the solution -- we set $P_0=1$ atm. The reaction plane is somewhere in the Stern layer, i.e. $x_\mathrm{rp} < x_2$. Because the pressure is constant for $x < x_2$ ($\rho_\mathrm{F}$ is zero), the pressure at the reaction plane is simply $P(x_\mathrm{rp})=P(x_2)$.

\begin{table}[]
\centering
\renewcommand{\arraystretch}{1.25}
\begin{tabular}{@{}ll@{}}
\toprule
\multicolumn{2}{l}{\textbf{General constants}}                                                                                                                                                     \\ \midrule
$k_\mathrm{B}$, Boltzmann constant                                                       & $1.38 \times 10^{-23}$ J/K                                                                              \\
$e$, elementary charge                                                                   & $1.6\times10^{-19}$ C                                                                                   \\
$N_\mathrm{A}$, Avogadro number                                                          & $6.02 \times 10^{23}$ /mol                                                                              \\
$\varepsilon_0$, vacuum permittivity                                                     & $8.85 \times 10^{-11}$ F/m                                                                              \\
\midrule
\multicolumn{2}{l}{\textbf{Electrolyte model}  }                                                                                                                                                   \\
\midrule
$T$, temperature                                                                         & 298 K                                                                                                   \\
$\beta$, inverse temperature                                                             & $(k_\mathrm{B} T)^{-1}$                                                                                 \\
$\varepsilon_\mathrm{w}$, rel. permittivity of pure water                                & 78.5                                                                                                    \\
$n_\mathrm{w}$, number density of pure water                                             & 55.5 M $\times N_\mathrm{A}$                                                                            \\
$n_\mathrm{max}$, lattice site density                                                   & $n_\mathrm{w}$                                                                                          \\
$i$ or $j$, species                                                                      & $\mathrm{H^+}, \mathrm{OH^-,} AM, X, \mathrm{H_2O}$                                                     \\
$\gamma_{AM}$, cation size factor                                                        & 3, 4, 5, 6; default: 4                                                                                  \\
$\gamma_{X}, \gamma_\mathrm{H^+}$, anion and proton size factor                          & 4                                                                                                       \\
$\gamma_\mathrm{OH^-}$, hydroxide ion size factor                                        & 3                                                                                                       \\
pH                                                                                       & 7, 13                                                                                                   \\
$\varepsilon_\infty$, optical permittivity                                               & $1.33^2$ \citep{iglivc2019differential}                                                                 \\
$p$, effective dipole moment of water                                                    & $\sqrt{\dfrac{3(\varepsilon_\mathrm{w} - \varepsilon_\infty)}{\beta n_\mathrm{w}}}$    \\
$d_\mathrm{H_2O}$, effective diameter of water                                           & $(n_\mathrm{w})^{-1/3}$                                                                                 \\
$d_{AM}$, effective diameter of cations                                                  & $(\gamma_{AM})^{1/3} d_\mathrm{H_2O} $                                                                  \\
$c_{AM}^\mathrm{b}$, bulk cation concentration                                           & 10, 100, 1000 mM                                                                                        \\
$c_\mathrm{H^+}^\mathrm{b}$, bulk proton concentration                                   & $10^{-\mathrm{pH}}$ M                                                                                   \\
$c_\mathrm{OH^-}^\mathrm{b}$, bulk hydroxide concentration                               & $10^{-14 + \mathrm{pH}}$ M                                                                              \\
$c_X^\mathrm{b}$, bulk anion concentration                                               & $c_\mathrm{H^+}^\mathrm{b} + c_{AM}^\mathrm{b} - c_\mathrm{OH^-}^\mathrm{b}$                            \\
$c_\mathrm{H_2O}^\mathrm{b}$, bulk water concentration                                   & $n_\mathrm{max} /N_\mathrm{A} - \sum_{i \in \mathrm{ions}} \gamma_i c_i^\mathrm{b}$      \\
$x_2$, distance of closest approach for cations                                          & $d_{AM}/2$                                                                                              \\
\midrule
\multicolumn{2}{l}{\textbf{Metal}  }                                                                                                                                                           \\
\midrule
$E_\mathrm{PZC,Au}$, point of zero charge of gold                                        & \qty{0.2}{\SHE} \citep{Maier2023}                                                 \\
\midrule
\multicolumn{2}{l}{\textbf{Insulator}  }                                                                                                                                                           \\
\midrule
$\bar n_\mathrm{sil}$, areal density of silanol groups                                   & $5\times 10^{18} \qty{}{m^{-2}}$ \citep{van1996general}                                                 \\
$K_\mathrm{a}$, acid dissociation constant for silanol                                   & $10^{-6}$ M \citep{van1996general}
\end{tabular}
\caption{Parameter table for the simulations of the electric double layer.}
\label{tab:parameter-table}

\end{table}

\section*{Acknowledgements}
The authors thank Dr. Jun Huang for his help with the numerical implementation of the double layer model. The authors gratefully acknowledge the support by Deutsche Forschungsgemeinschaft (DFG, German Research Foundation) through 'e-conversion Cluster of Excellence', EXC 2089/1-390776260, and through TUM International Graduate School of Science and Engineering (IGSSE), GSC 81-24184165, and the support by the Bavarian State Ministry of Science and the Arts within the Collaborative Research Network 'Solar Technologies go Hybrid (SolTech)'.

\newpage
\printbibliography

\newpage
\appendix
\section{Appendix}
\label{app:1}

\subsection{Geometric properties of electrodes}\label{app:geometric-properties}

The geometric properties of the metal array on the substrate surface is determined by the geometric properties of the stamp used in the LO-NIL fabrication process, cf. to \cite{Maier2020, Golibrzuch2022} for more information. The geometric properties of the stamps used for producing the arrays consisting of nanostructures in different sizes is summarized in Table \ref{tab:exp_geometric-properties_structures}.

\begin{table}[h!]
	\begin{center}
		\begin{tabular}{r|l|l|l}
			\toprule
			Size $d$ / nm & Structure shape & Pitch ratio $d/p$ & Filling factor \\
			\midrule
			1400 & Squares & 0.47 & 22 \% \\
			350 & Squares & 0.50 & 25 \% \\
			200 & Squares & 0.46 & 21 \% \\
			75 & Circles & 0.50 & 20 \% \\
			\bottomrule
		\end{tabular}
	\end{center}
	\caption{Relevant geometric parameters of the metal arrays present on the structured electrodes investigated in this work. The size $d$ labels the size of single structures in the array and the pitch $p$ the center-to-center distance between individual structures. The filling factor gives the geometric share of metal covered to non-covered surface.}
	\label{tab:exp_geometric-properties_structures}
\end{table}

\subsection{Influence of the rotation speed on the HER activity}\label{app:rotation-speed}
The measurements shown in this work are conducted without electrode rotation. Thus, the influence of rotation on the HER activity of the electrodes is elaborated here. The setup used is a custom-built recessed rotating disk electrode setup (recess depth: $\qty{1.6}{\milli\meter}$, diameter of exposed electrode surface: $\qty{8.5}{\milli\meter}$), which turned out to behave as an ideal RDE setup for the electrodes considered here (maximum diameter of active surface area: \qty{6.0}{\milli\meter}). More information regarding the ReRDE setup can be found in \cite{Maier2023}.

\begin{figure}[h!]
	\centering
	\includegraphics[width=0.49\textwidth]{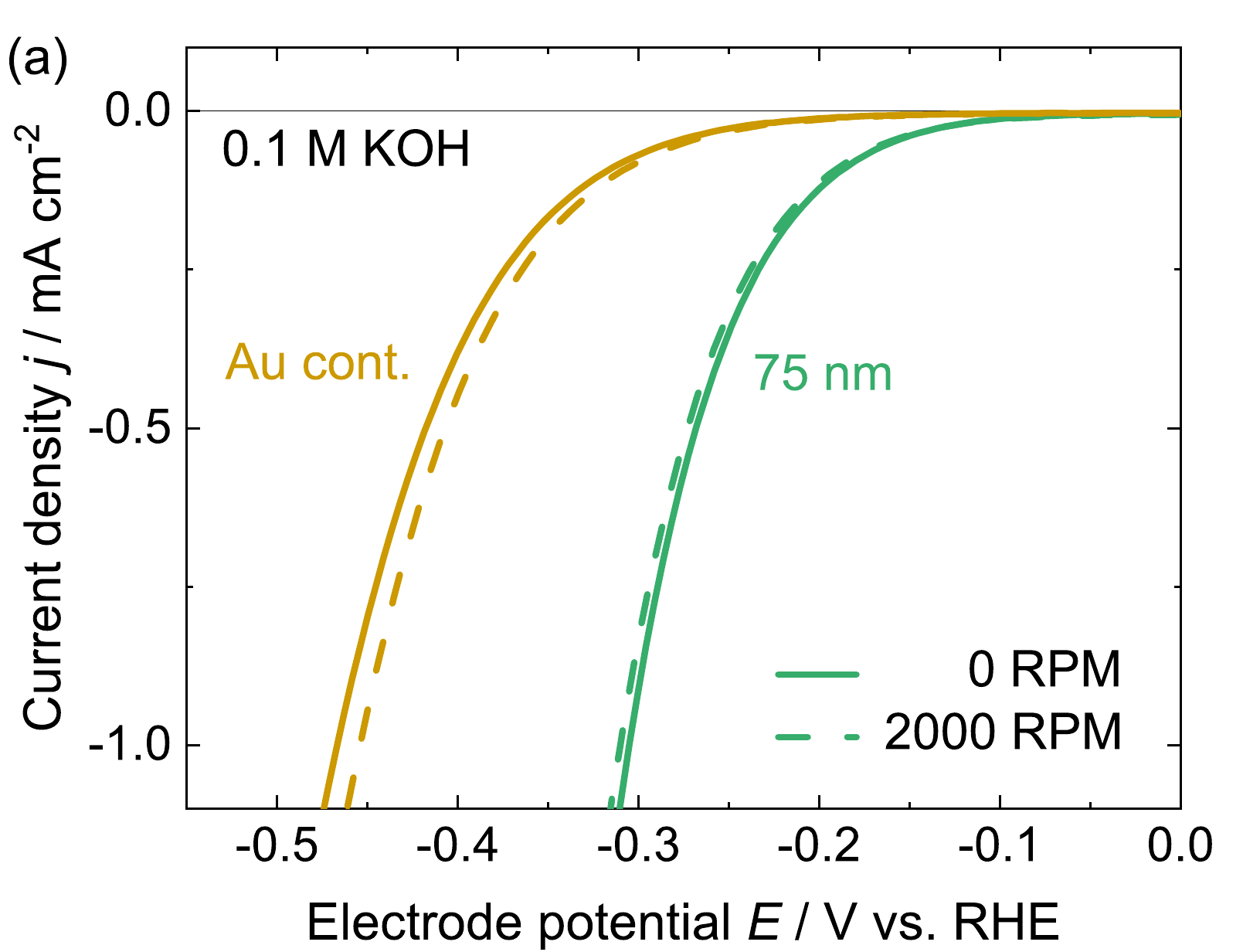}
	\includegraphics[width=0.49\textwidth]{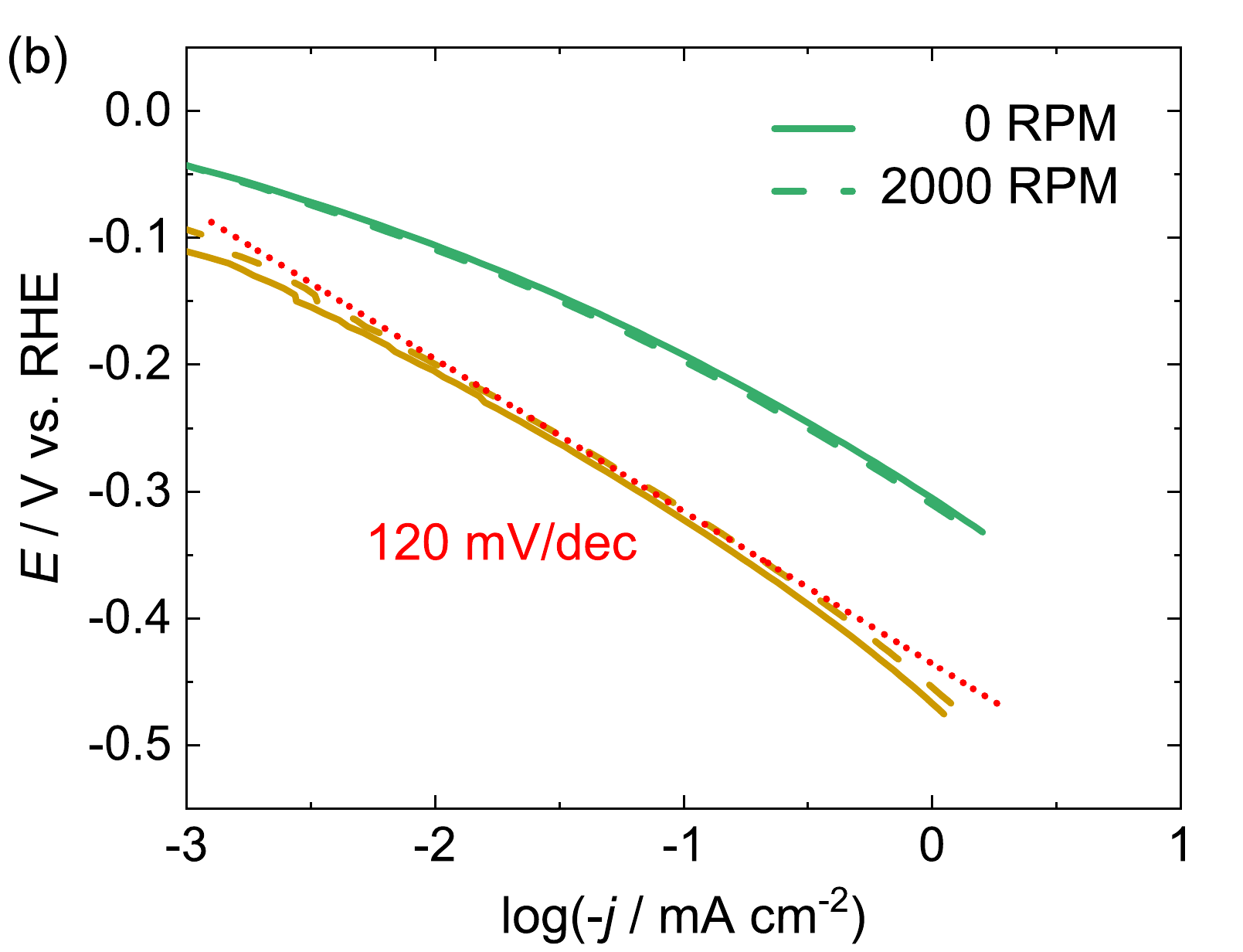}
	\caption{(a) Comparison of the LSVs of a continuous Au layer electrode (yellow lines) and a nanostructured electrode with 75 nm Au structures (greed lines) in $\qty{0.1}{\molar} ~\ce{KOH}$ (pH 13) without rotation (i.e. rotation speed 0, solid lines) and with rotation (i.e. rotation speed 2000 RPM, dashed lines). (b) Corresponding Tafel plots after subtraction of the non-faradaic current at $\qty{0}{\RHE}$. The dotted red line is given as a guide to the eye.}
	\label{fig:rotation-speed}
\end{figure}

Figure \ref{fig:rotation-speed}(a) shows the comparison between the LSVs of the continuous Au layer electrode (yellow lines) and a nanostructured electrode with 75 nm Au structures (green lines) in $\qty{0.1}{\molar} ~\ce{KOH}$ (pH 13) without rotation (i.e. rotation speed at 0 RPM, solid lines) and with rotation (i.e. rotation speed at 2000 RPM, dashed lines). It becomes evident that rotation does not alter the observed behavior significantly. The HER activity of the continuous Au layer appears to become slightly higher with rotation, while the activity of the nanostructured electrode does not change considerably. Figure \ref{fig:rotation-speed}(b) shows the corresponding Tafel plots after subtraction of the non-faradaic current at $\qty{0}{\RHE}$. It is evident that also the slope of the Tafel plots does not change significantly when rotating the electrode.

\subsection{CVs of the evaporated Au film}\label{app:CVs-Au-film}

\begin{figure}[!htb]
	\centering
	\includegraphics[width=0.49\textwidth]{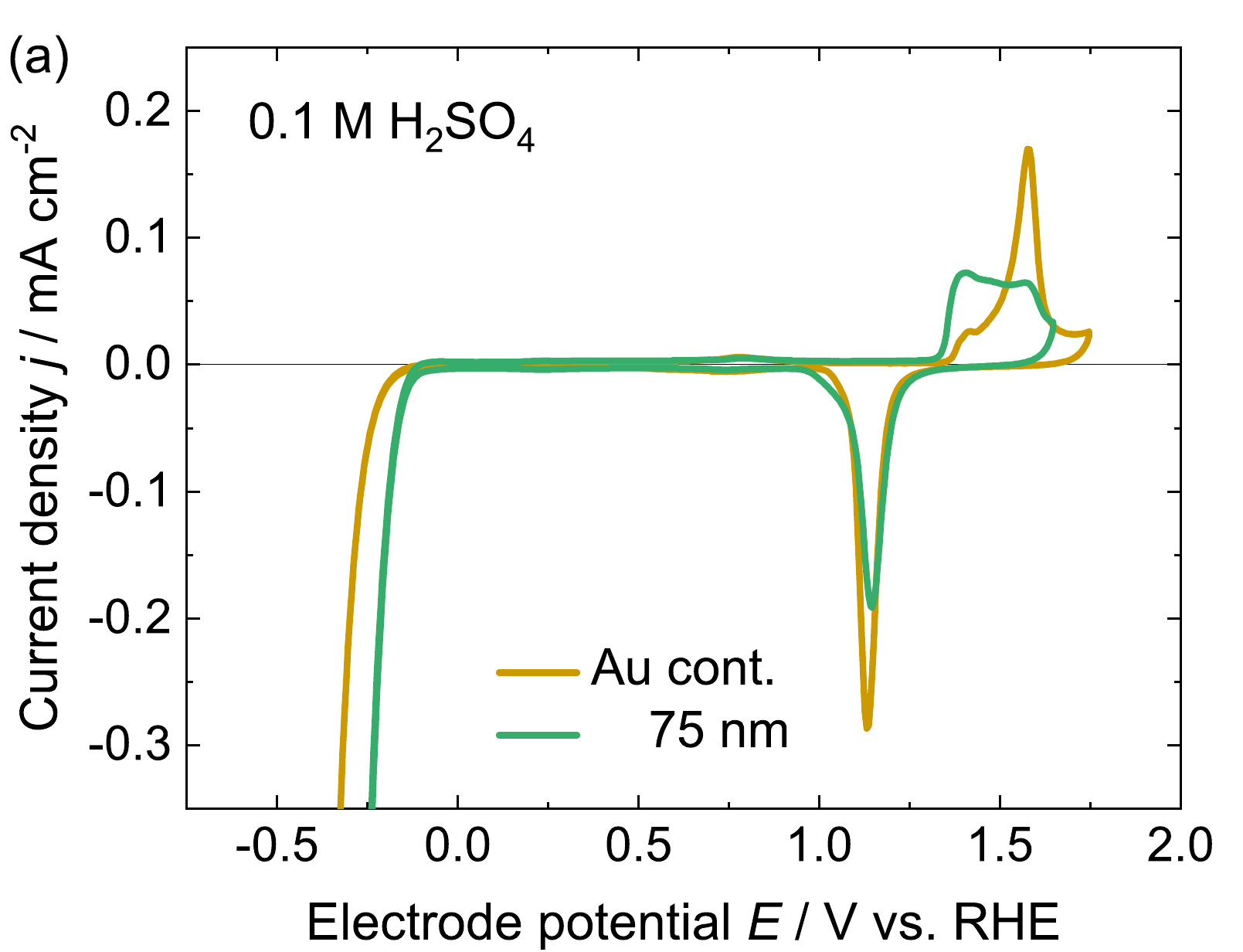}
	\includegraphics[width=0.49\textwidth]{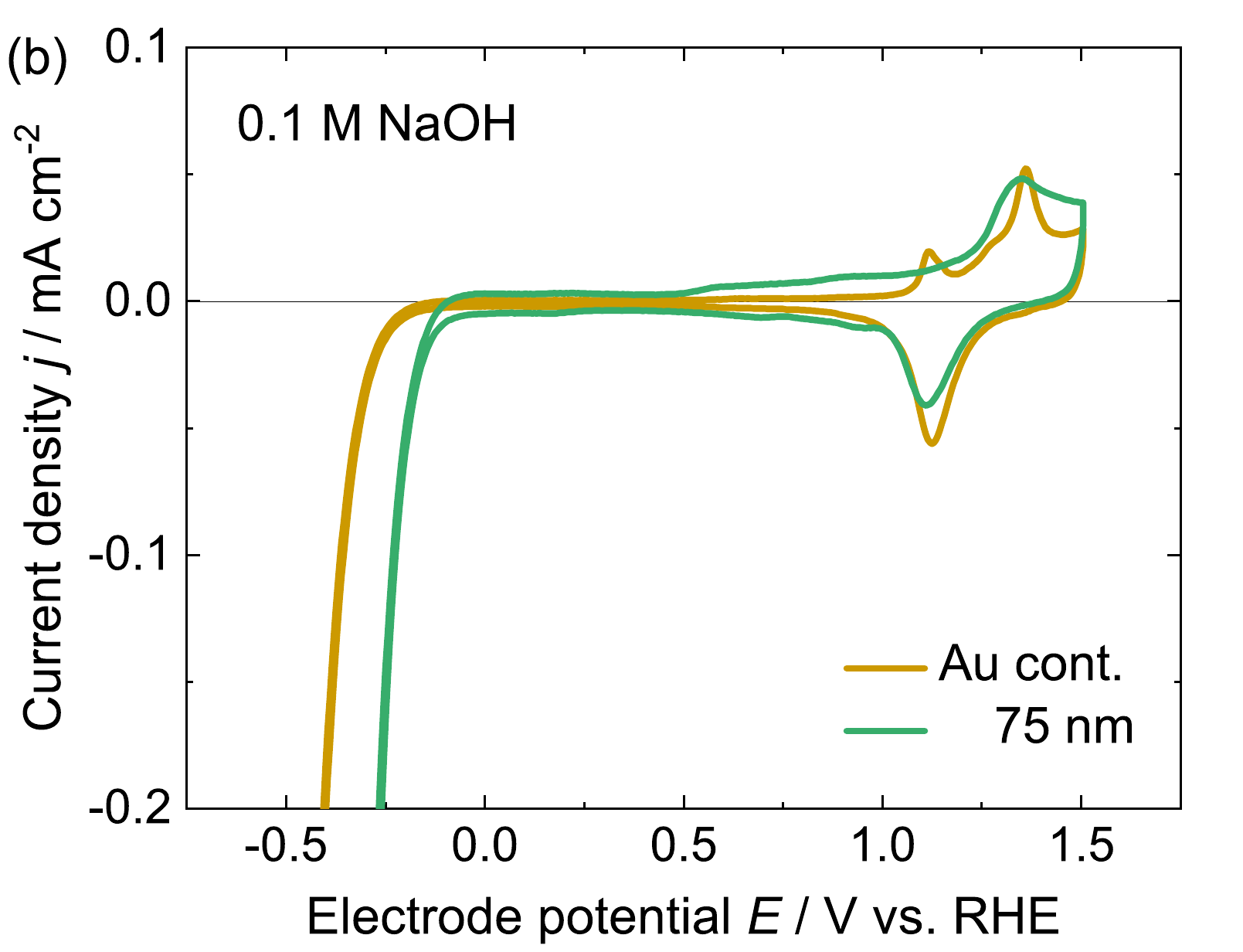}
	\caption{Exemplary CVs of the evaporated Au film (continuous layer electrode, yellow) and a nanostructured electrode with 75 nm structure size (green) in (a) $\qty{0.1}{\molar}~\ce{H2SO4}$ (pH 1) and (b) in $\qty{0.1}{\molar}~\ce{NaOH}$/Ar sat.'d (pH 13) electrolyte. Scan rate: $\qty{50}{\milli\volt\per\second}$.}
	\label{fig:Au-CVs}
\end{figure}

Figure \ref{fig:Au-CVs} shows exemplary CVs of the evaporated Au-film on silicon-based substrate, i.e. a typical continuous Au layer electrode (yellow), and exemplary CVs of the nanostructured electrode with 75 nm structure size (green) for comparison, in (a) $\qty{0.1}{\molar}~\ce{H2SO4}$ and in (b) $\qty{0.1}{\molar}~\ce{NaOH}$.

\subsection{Influence of cationic species on the HER activity in neutral electrolyte}\label{app:cation-species-neutral}

\begin{figure}[h!]
	\centering
	\includegraphics[width=0.55\textwidth]{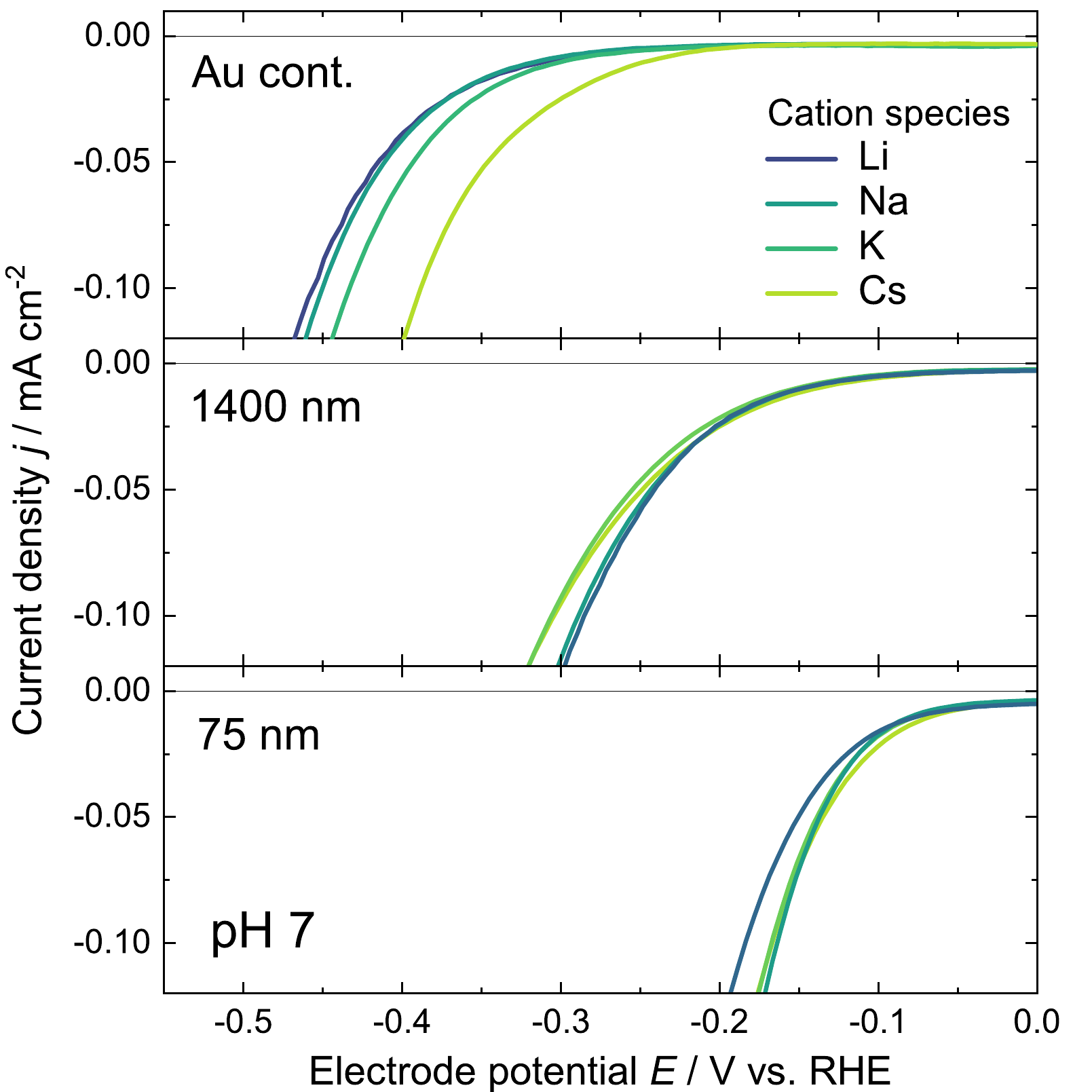}
	\caption{Dependence of the HER activity on the cationic species in neutral electrolyte. The electrolytes consist of $\qty{0.18}{\molar}~AM\ce{OH}$ ($AM = \ce{Li}, \ce{Na}, \ce{K}, \ce{Cs}$) and $\qty{0.12}{\molar}~\ce{H2PO4-}/\ce{HPO4^{2-}}$ (pH 7). Top: A continuous gold layer. Middle: A nanostrutured electrode with $\qty{1400}{\nano\meter}$ metal island diameter. Bottom: A nanostructured electrode with $\qty{75}{\nano\meter}$ metal island diameter. Scan rate: $\qty{50}{\milli\volt\per\second}$.}
	\label{fig:cation-species_neutral}
\end{figure}

Figure \ref{fig:cation-species_neutral} shows LSVs of the three different electrode systems in a buffered neutral electrolyte (pH 7). It is evident that the HER improvement for smaller structures is seen for all cationic species. This behavior is similar to the behavior in alkaline electrolyte shown in the main text. Contrary to the situation in alkaline electrolyte, the spreading between LSV scans (i.e. the distance between the curves at a certain current density) at different cationic species is rather small for nanostructured electrodes.

A possible explanation for this behavior can be given on the basis of the proposed mechanism: Since the electrode surface is less polarized at the same potential in RHE scale in neutral electrolyte compared to alkaline electrolyte, the net concentration of cations is supposed to be lower in the prior case. A lower net concentration of cations is non-optimal for the standard HER process at a continuous Au surface, but rather irrelevant for the mechanism at the metal/insulator interface (which leads to the great distance between Au cont. and 75 nm in the figure). In neutral electrolyte, the net concentration of cations in the double layer above the insulator surface is rather low. As in the nanostructured case the reaction speed is mainly dominated by the charge transport through the double layer, it is supposed to be not as strongly dependent on the cation identity as it is in the alkaline case, in which a higher concentration of cations in the double layer is present. This would lead to the smaller spreading of curves for nanostructured electrodes in neutral electrolyte, which is observed.

\subsection{Composition of electrolytes with different cation concentration}\label{app:composition-cation-concentration}

Table \ref{tab:res_nsi_cation_concentration} shows the composition of the electrolytes considered when comparing the HER acitivity of the nanostructured electrode at various concentrated electrolytes (cf. Figure \ref{fig:cation-concentration}). The electrolytes are buffered using a phosphate buffer with an approximate similar buffer strength (except for the most dilute electrolyte). Note that the electrolytes are adjusted to have the same pH value of 7, which may lead to slightly differing actual buffer concentration, as given here in the table. The various cation concentrations are adjusted by adding various amounts of salt (here $\ce{K2SO4}$).

\begin{table}[h!]
	\centering
	\begin{tabular} {r||r|r|r|r}
		\toprule
		$\ce{K+}$ Electrolyte &  \multicolumn{3}{c|}{Approx. ion concentrations [mM]} &  Approx. ionic strength [mM] \\
		& [$\ce{K+}$] & [$\ce{SO4^2-}$] & [$\ce{H2PO4-}/\ce{HPO4^2-}$] & \\\midrule
		10 mM	& 10 	& 0 	& 6.7	& 13	\\
		30 mM	& 30 	& 6		& 12	& 42	\\
		100 mM 	& 100 	& 41	& 12	& 147	\\
		300 mM 	& 300 	& 141	& 12	& 447	\\
		1000 mM & 1000 	& 491	& 12	& 1497	\\
		\bottomrule
	\end{tabular}
	\caption{Overview of the composition (approximate ion concentrations) and the resulting ionic strengths of the respective electrolytes used in this section. All electrolytes are adjusted to have a similar pH value of 7.}
	\label{tab:res_nsi_cation_concentration}
\end{table}

\subsection{Cu deposition and stripping}\label{app:cu-deposition-stripping}

\begin{figure}[h!]
	\centering
	\includegraphics[width=0.49\textwidth]{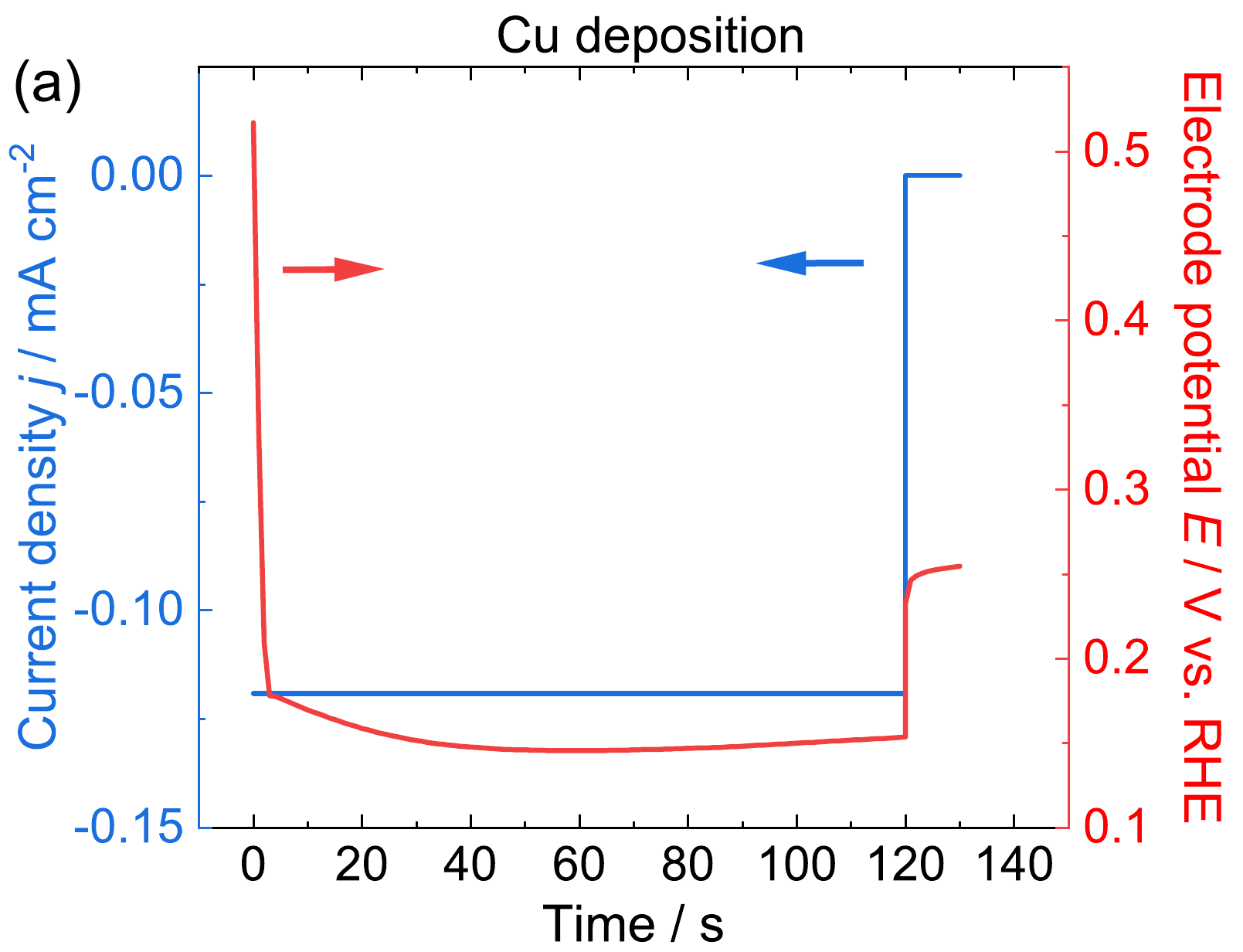}
	\includegraphics[width=0.49\textwidth]{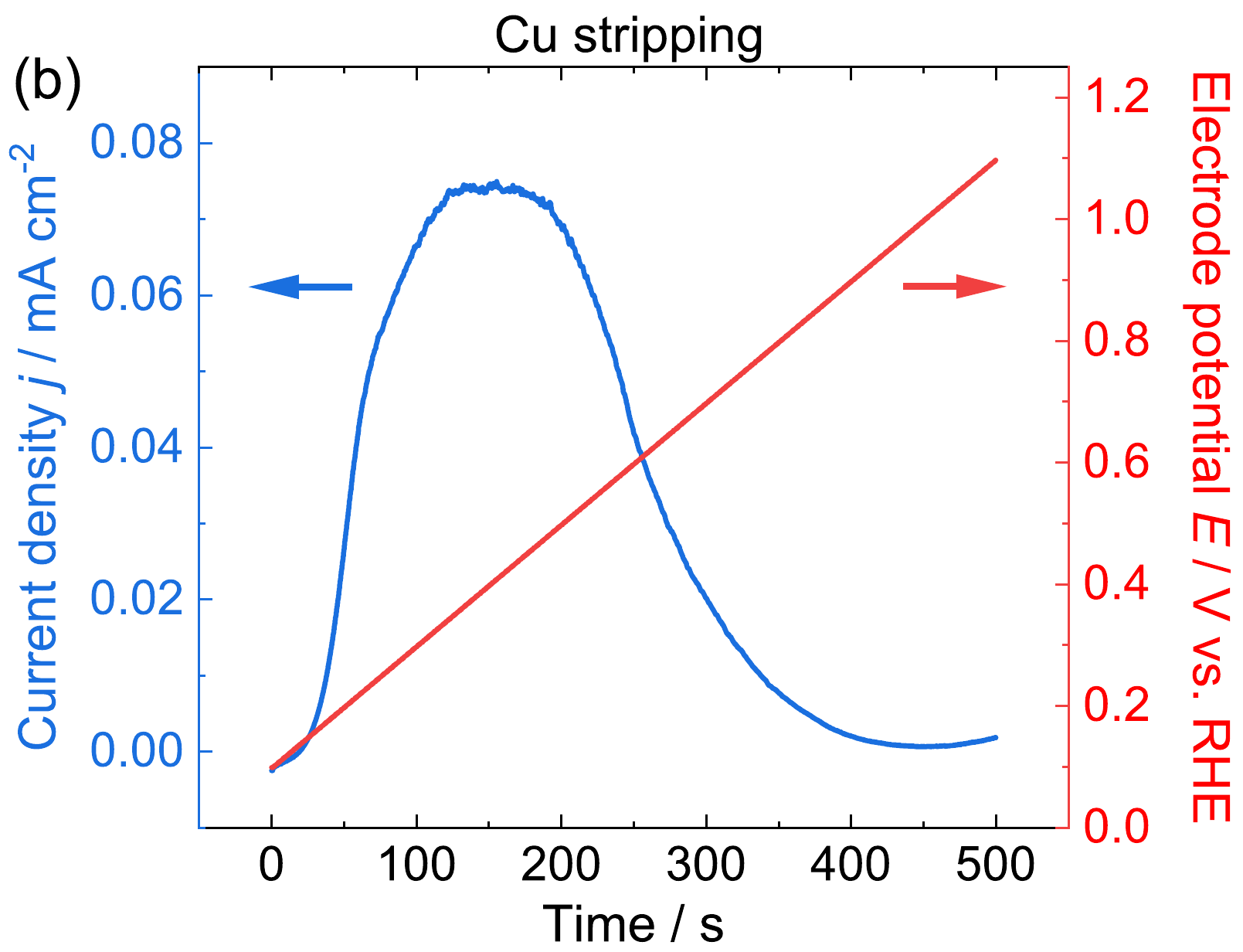}
	\caption{(a) Exemplary galvanostatic copper deposition on a nanostructured gold electrode with structure size of \qty{1400}{\nano\meter}. Blue: Current density as a function of time. Red: Applied electrode potential as a function of time. The deposition is conducted in $\qty{1}{\milli\molar} ~\ce{CuSO4}$ + $\qty{0.1}{\molar} ~\ce{H2SO4}$/Ar sat.'d (pH 1) at $\qty{-0.12}{\milli\ampere\per\square\centi\meter}$. (b) Stripping of the Cu layer in $\qty{0.18}{\molar}~\ce{NaOH}$ + $\qty{0.12}{\molar}~\ce{H2PO4^-}/\ce{HPO4^2-}$, Ar sat.'d (pH 7) with a potential scan rate of $\qty{2}{\milli\volt\per\second}$. }
	\label{fig:copper-deposition-stripping}
\end{figure}

The copper layer is deposited onto the Au layer of the electrodes by electrochemical copper deposition. The deposition is conducted galvanostatically and the amount of copper deposited is chosen to be approx. 30 monolayers. We consider a specific charge of $\qty{460}{\micro\coulomb\per\square\centi\meter}$ \cite{Hachiya1991} for one monolayer of Cu atoms on a Au(111) surface. This rather high copper coverage is chosen to assure that the catalytic behavior of the surface is fully determined by (bulk) copper properties. The deposition electrolyte is chosen to be $\qty{1}{\milli\molar} ~\ce{CuSO4}$ + $\qty{0.1}{\molar} ~\ce{H2SO4}$ (pH 1). Figure \ref{fig:copper-deposition-stripping}(a) shows an exemplary copper deposition experiment onto a $\qty{1400}{\nano\meter}$ electrode.

After deposition, the electrolyte is exchanged to the 'working electrolyte' consisting of $\qty{0.18}{\molar}~\ce{NaOH}$ + $\qty{0.12}{\molar}~\ce{H2PO4^-}/\ce{HPO4^2-}$, Ar sat.'d (pH 7). During the exchange process a continuous flow of Ar gas into the cell is realized to avoid oxidation of the Cu surface and the electrode potential is controlled. The HER activities of the electrodes are measured in this electrolyte.

Subsequent to the measurement process, the copper layer is stripped again to ascertain the amount of copper lost during the exchange process and the activity measurement. A typical stripping experiment is shown in Figure \ref{fig:copper-deposition-stripping}(b). By comparing the charge transferred during deposition and stripping, it turned out that almost no copper is lost during the exchange of the electrolyte and subsequent HER measurements.

\subsection{CVs of the evaporated Pt film}\label{app:CVs-Pt-film}
\begin{figure}[h!]
	\centering
	\includegraphics[width=0.49\textwidth]{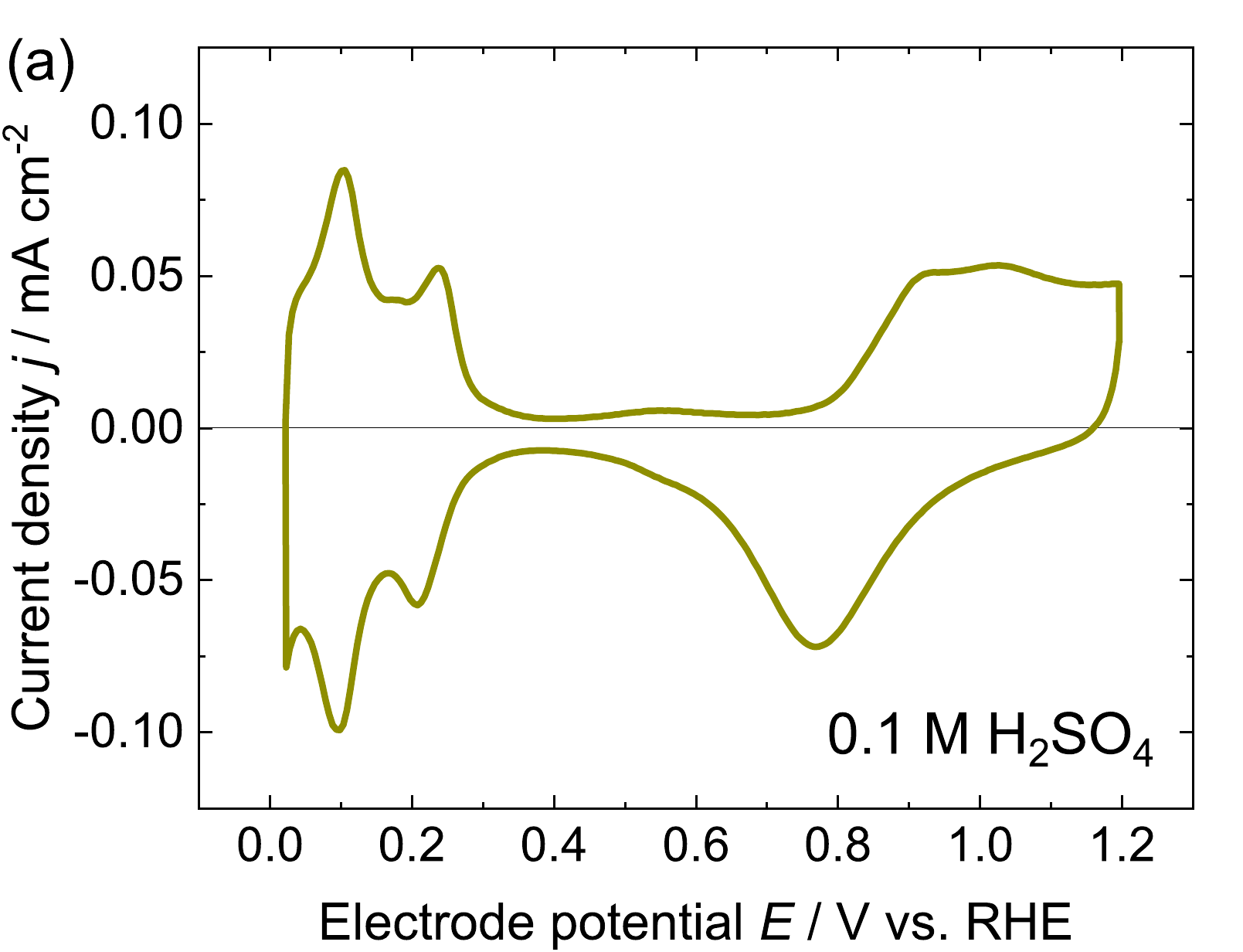}
	\includegraphics[width=0.49\textwidth]{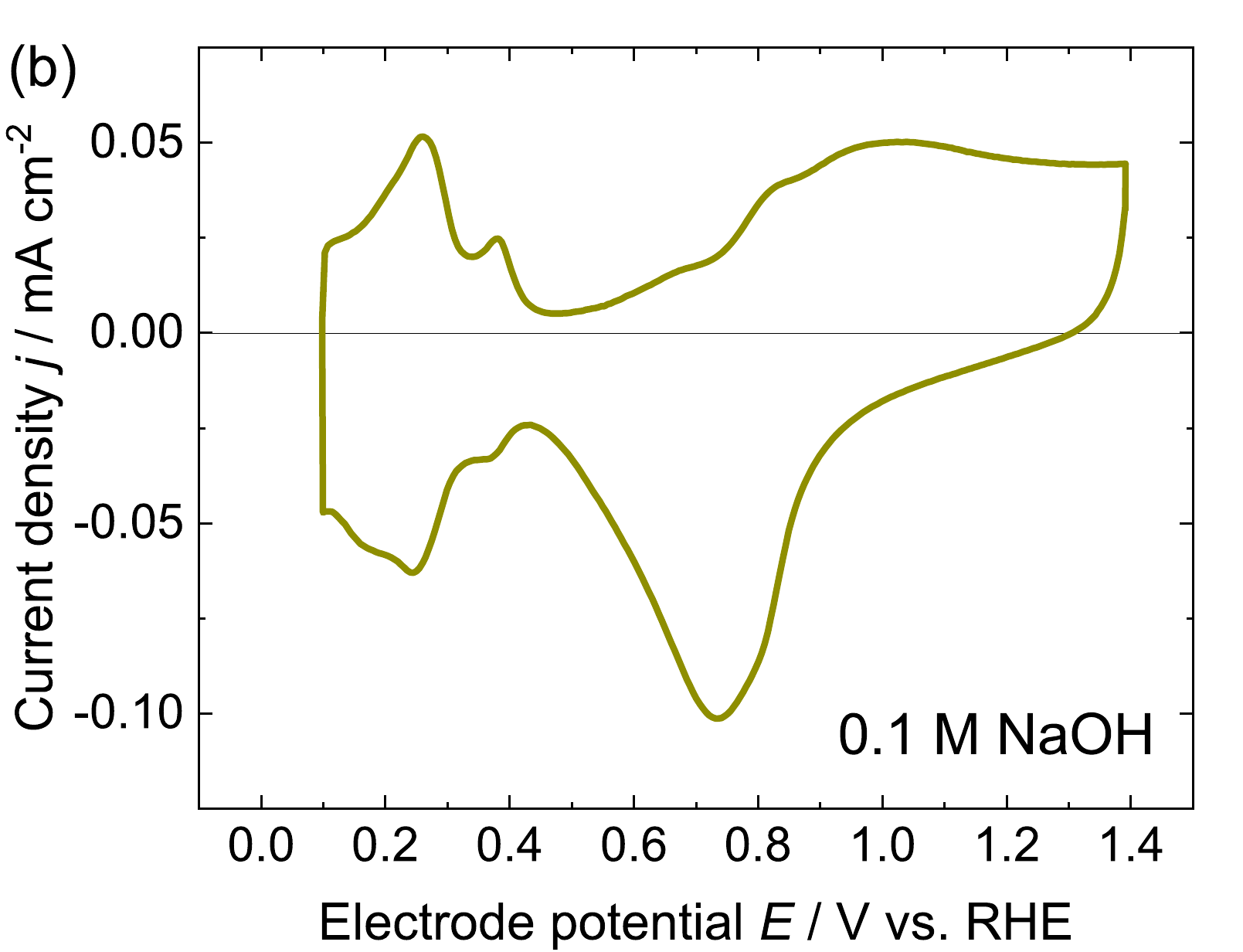}
	\caption{CVs of the evaporated Pt film (continuous layer electrode) in acidic and alkaline electrolyte: (a) In $\qty{0.1}{\molar}~\ce{H2SO4}$/Ar sat.'d, (b) in $\qty{0.1}{\molar}~\ce{NaOH}$/Ar sat.'d. Scan rate: $\qty{50}{\milli\volt\per\second}$.}
	\label{fig:Pt-CVs}
\end{figure}

The electrodes consisting of a Pt surface (continuous layer electrode or nanostructured electrodes) are fabricated by the same process as the corresponding, i.e., , i.e. only the nature of the evaporated metal was changed from Au to Pt. An exemplary CV of the evaporated Pt film (continuous layer electrode) showing the quality of the film is shown in Figure \ref{fig:Pt-CVs}. All relevant features present for Pt(pc) systems known from the literature \cite{Rheinlaender2014, Sheng2010} are visible in the data.

\subsection{AuOx reduction charge and roughness factor of Au films}\label{app:auox-reduction_utp}

\begin{figure}[h!]
	\centering
	\includegraphics[width=0.49\textwidth]{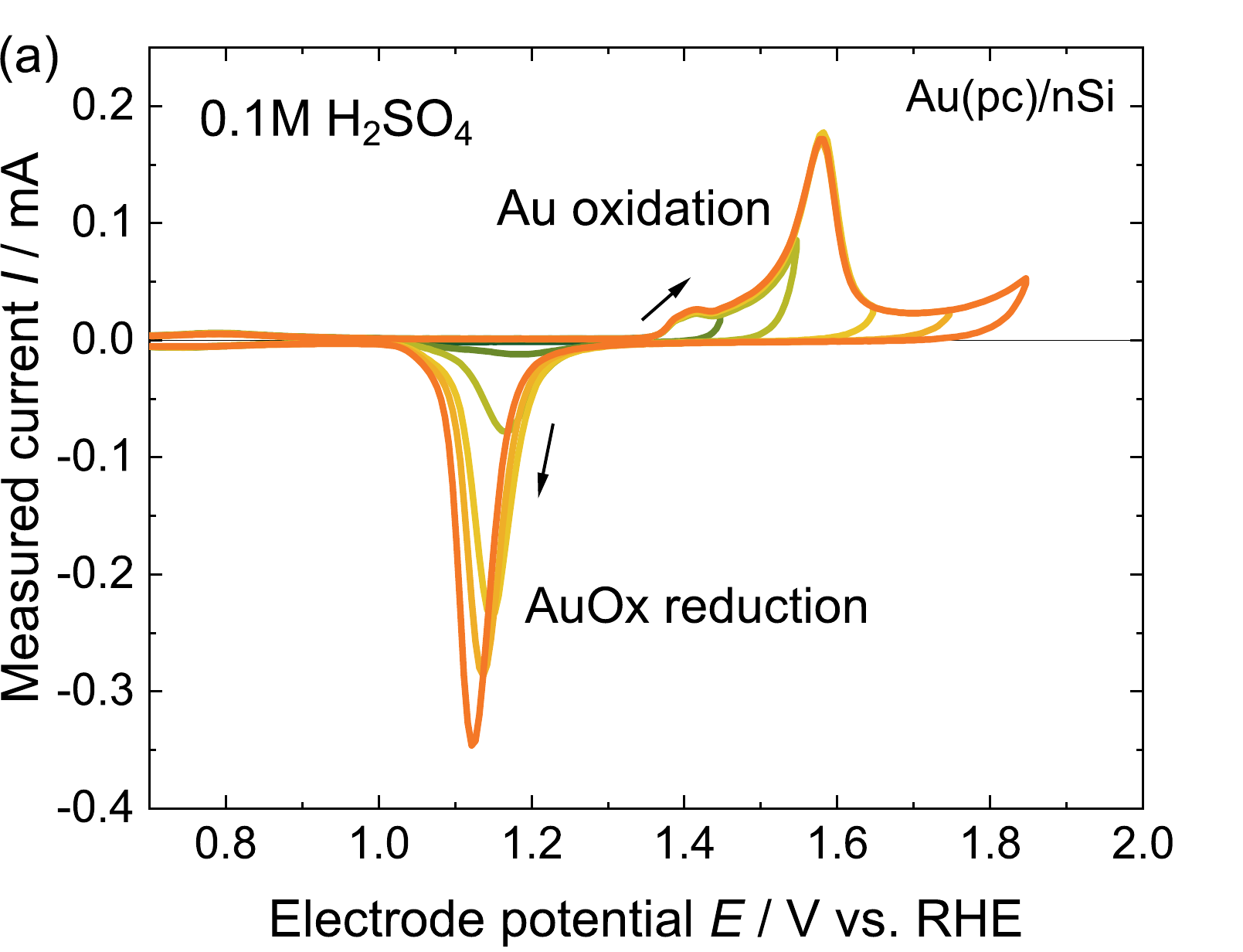}
	\includegraphics[width=0.49\textwidth]{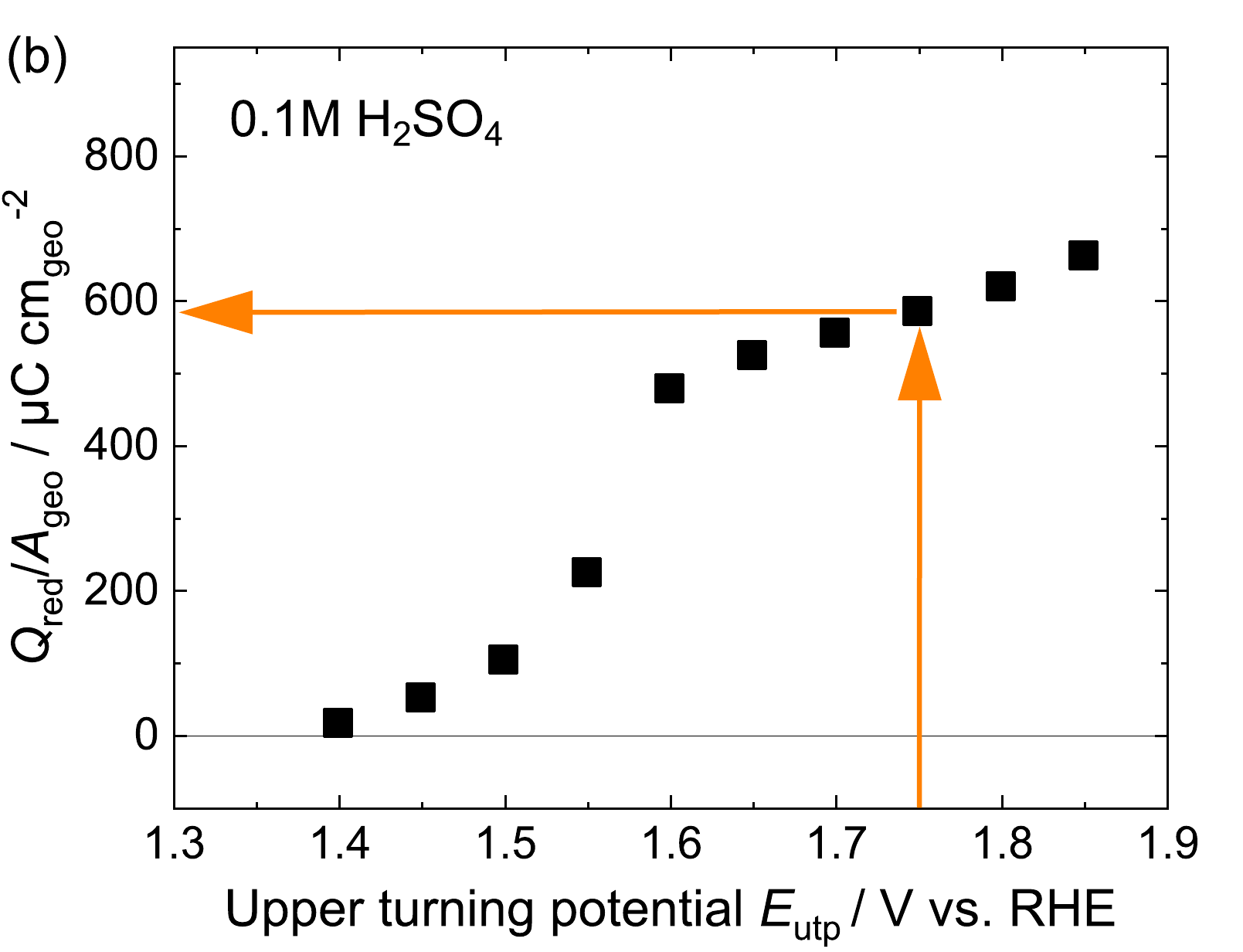}
	\caption{(a) CVs of an evaporated Au film (continuous layer electrode) in $\qty{0.1}{\molar}~\ce{H2SO4}$/Ar sat.'d for various upper turning potential potential (utp). (b) Charge transferred during the AuOx reduction process normalized to the geometric gold area $A_\mathrm{geo}$ as a function of the upper turning potential.}
	\label{fig:Au-area-utp}
\end{figure}

The electrochemical active surface area of Au-based electrodes is determined from OH-desorption (i.e. AuOx reduction). Figure \ref{fig:Au-area-utp}(a) shows CVs of a typical Au(pc) electrode (a continuous Au layer electrode as evaporated) for various upper turning potentials in $\qty{0.1}{\molar}~\ce{H2SO4}$. Figure \ref{fig:Au-area-utp}(b) shows the transferred charge during the AuOx reduction process normalized to the geometric Au area (macroscopic geometric size of the evaporated Au film in contact with the electrolyte) as a function of the upper turning potential $E_\mathrm{utp}$. Assuming a 'full coverage' of the Au surface at $E_\mathrm{utp}= \qty{1.75}{\RHE}$ results in a roughness factor of $RF \approx 1.5$, when dividing the corresponding reduction charge, see orange arrows in the Figure \ref{fig:Au-area-utp}(b), by the specific surface charge value for Au surfaces of $\qty{390}{\micro\coulomb\per\square\centi\meter}$ \cite{Trasatti1991, BardFaulkner2001}.

\subsection{Numerical solution of the double-layer model} \label{ssec:model-numerical-solution}
\subsubsection{Dimensionless quantities}
To solve equation \eqref{eq:mpbe} numerically, we first nondimensionalize the equation. To this end we use an inverse length scale denoted as $\kappa$,
\begin{equation}
    \kappa = \sqrt{\frac{2 \beta n_\mathrm{max} e^2}{\varepsilon_0\varepsilon_\mathrm{w}}}
\end{equation}
with $n_\mathrm{max}=(d_\mathrm{H_2O})^{-3}$ the lattice site density and $\varepsilon_\mathrm{w}\approx78.5$ the bulk relative permittivity of water. $\kappa$ can be interpreted as the inverse Debye length \citep{Schmickler2010} for the case that the lattice is tightly packed with ions of size $(d_\mathrm{H_2O})^3$, a case that yields a large electric field. Using $\kappa$ we define the dimensionless potential $y_1$, electric field $y_2$, spatial coordinate $\zeta$, rescaled permittivity $\tilde{\varepsilon}$ and dimensionless dipole moment magnitude $\tilde{p}$ as
\begin{equation}
\begin{aligned}[c]
    y_1 = \beta e\phi,
\end{aligned}
\qquad
\begin{aligned}[c]
y_2 = \frac{\beta e }{\kappa} \frac{\partial \phi}{\partial x},
\end{aligned}
\qquad
\begin{aligned}[c]
\zeta = \kappa (x-x_2),
\end{aligned}
\qquad
\begin{aligned}[c]
    \tilde{\varepsilon} = \frac{\varepsilon}{\varepsilon_\mathrm{w}},
\end{aligned}
\qquad
\begin{aligned}[c]
    \tilde{p}=\frac{\kappa p}{e}.
\end{aligned}
\end{equation}
where $p$ is the magnitude of the effective dipole moment of water. With the above quantities, equation \eqref{eq:mpbe} can be rewritten to
\begin{equation} \label{eq:nondim-pbe}
    \frac{\partial}{\partial\zeta} (\tilde{\varepsilon} y_2) = - \frac12 \sum_i \frac{z_i n_i}{n_\mathrm{max}}.
\end{equation}
The number densities are already defined in terms of dimensionless quantities in equation \eqref{eq:model-number-densities} (upon division by $n_\mathrm{max}$). The Boltzmann factors $\Theta_i$ expressed in terms of dimensionless quantities are
\begin{equation}
    \Theta_i = \begin{dcases}
        \exp(-z_i y_1), & i \in \{\mathrm{H^+, OH^-, +, - \};}\\
        \frac{\sinh{\tilde{p}y_2}}{\tilde{p}y_2}, &\text{for $i=\mathrm{H_2O}$.}
    \end{dcases}
\end{equation}
Finally, the expression for the permittivity of \cite{huang2021cation} is rewritten in terms of dimensionless quantities as
\begin{equation}
    \tilde{\varepsilon} = \frac{\varepsilon_\infty}{\varepsilon_\mathrm{w}} + \tilde{p}^2 \frac{n_\mathrm{H_2O}}{2n_\mathrm{max}} \frac{\mathcal{L}(\tilde{p}y_2)}{\tilde{p}y_2}
\end{equation}
with $\mathcal{L}(x)=\coth(x)-1/x$ the Langevin function.

\subsubsection{Numerical solution}
For the numerical solution of the boundary value problem we use SciPy's solve\_bvp function \citep{2020SciPy-NMeth}. To solve equation \eqref{eq:nondim-pbe} with solve\_bvp we must rewrite the equation as a system of two first order differential equations. To this end we define
\begin{align}
    F_1 &= -\sum_i z_i \frac{n_i}{2n_\mathrm{max}} \\
    F_2 &= - \tilde{p} y_2 \mathcal{L}(\tilde{p}y_2) \frac{n_\mathrm{H_2O}}{2 n_\mathrm{max}} \sum_i z_i \gamma_i \frac{n_i}{n_\mathrm{max}} \\
    G_1 &= \frac{\varepsilon_\infty}{\varepsilon_\mathrm{w}} \\
    G_2 &= \tilde{p}^2 \frac{n_\mathrm{H_2O}}{2n_\mathrm{max}} \mathcal{L}'(\tilde{p}y_2) \\
    G_3 &= \tilde{p}^2 \mathcal{L}^2(\tilde{p} y_2) \frac{n_\mathrm{H_2O}}{2n_\mathrm{max}} \left(1-\frac{n_\mathrm{H_2O}}{n_\mathrm{max}}\right)
\end{align}
where $\mathcal{L}'(x)=\mathrm{d}\mathcal{L}(x)/\mathrm{d} x$ and $\mathcal{L}^2(x)=(\mathcal{L}(x))^2$. The system of first-order differential equations is then
\begin{equation}
    \begin{dcases}
        \frac{\partial y_1}{\partial \zeta} &= y_2 \\
        \frac{\partial y_2}{\partial \zeta} &= \frac{F_1 + F_2}{G_1 + G_2 + G_3}.
    \end{dcases}
\end{equation}
The nondimensionalized boundary conditions for the metal, see equation \eqref{eq:metalbc}, are
\begin{equation}
    \begin{dcases}
        y_1(\zeta_\mathrm{end})&=0 \\
        y_1(0) &= \beta e \phi_0 + y_2(0) \kappa x_2
    \end{dcases}
\end{equation}
where we chose $\zeta_\mathrm{end}$ to correspond to $x= 100 \text{ nm}$, i.e. far away in the electrolyte as compared to the double layer thickness. As initial $\zeta$-axis we chose a logarithmically spaced axis so that there are more points in the double layer region and less points in the bulk electrolyte. The boundary condition for the insulator is nondimensionalized to
\begin{equation}
    \frac{\varepsilon (0)}{\varepsilon_\mathrm{w}} y_2(0) = \frac{\kappa \bar n_\mathrm{sil}}{2n_\mathrm{max}} \frac{K_\mathrm{a}}{K_\mathrm{a} + c_\mathrm{H^+}^\mathrm{b} \exp (-y_1(0) + \kappa x_2 y_2(0))}.
\end{equation}


To obtain a solution for a metal surface at arbitrary $\phi_0$, we first solve at the PZC ($\phi_0=0$) and then sweep to the desired potential in steps of $0.01$ V, each time using the solution as initial condition for the next iteration. To solve the insulator-electrolyte interface at arbitrary pH, we start at low pH, e.g. pH 2, where the surface charge is low, and sweep to the desired pH in steps of 0.25. From the solutions $y_1, y_2$, the relevant physical quantities $\phi$, $\mathcal{E}$, $c_i=n_i/N_\mathrm{A}$, and $\varepsilon$ were calculated and used as described in the main text.


\end{document}